\documentclass[a4paper,pra,showpacs,superscriptaddress,twocolumn]{revtex4}

\usepackage{amsfonts}
\usepackage{amssymb}
\usepackage{amsmath}
\usepackage{calc}
\usepackage{graphicx}
\usepackage{array}
\usepackage{bm}
\usepackage{delarray}
\usepackage{ifthen}

\newcommand{\sqrtfrac}[2]{\ifthenelse{\equal{#1}{1}}{\frac{\,\,1}{\sqrt{#2}}}{\frac{\sqrt{#1}}{\sqrt{#2}}}}

\def\advecb{{\bm\Phi}}
\def\advec{{\Phi}}
\def\hdots{\cdots}
\def\t{(t)}
\def\eigenv{\varepsilon}

\def\bwsbox#1{}

\def\zkbox#1{}
\def\bwsbox#1{\fbox{BWS: #1}}  
\def\zkbox#1{\fbox{ZK: #1}}  

\begin{document}
\title{Stimulated    Raman    Adiabatic    Passage   (STIRAP)    Among
  Degenerate-Level Manifolds}

\author{Z. Kis}
\affiliation{H.A.S. Research Institute for Solid State Physics and
Optics, H-1525 Budapest, P.O.Box 49, Hungary}
\affiliation{Fachbereich Physik der  Universit\"at Kaiserslautern,
67653 Kaiserslautern, Germany}
\author{A. Karpati}
\affiliation{H.A.S. Research Institute for Solid State Physics and
Optics, H-1525 Budapest, P.O.Box 49, Hungary}
\author{B. W. Shore}
\affiliation{618 Escondido Cir., Livermore CA 94550, USA}
\affiliation{Fachbereich Physik der  Universit\"at Kaiserslautern,
67653 Kaiserslautern, Germany}
\author{N. V. Vitanov}
\affiliation{Department of Physics, Sofia University, James Boucher 5
blvd., 1164 Sofia, Bulgaria}
\altaffiliation{Institute of Solid State Physics,
   Bulgarian Academy of Sciences, Tsarigradsko chauss\'{e}e 72, 1784
Sofia, Bulgaria}
\affiliation{Fachbereich Physik der  Universit\"at Kaiserslautern,
67653 Kaiserslautern, Germany}

\begin{abstract}
  We  examine the  conditions  needed to  accomplish stimulated  Raman
  adiabatic passage (STIRAP) when the  three levels ($g$, $e$ and $f$)
  are  degenerate,  with   arbitrary  couplings  contributing  to  the
  pump-pulse  interaction   ($g$  -  $e$)  and   to  the  Stokes-pulse
  interaction ($e$-$f$).   We show that in general  a {\em sufficient}
  condition  for  complete population  removal  from  the  $g$ set  of
  degenerate  states for arbitrary,  pure or  mixed, initial  state is
  that the  degeneracies should not  decrease along the  sequence $g$,
  $e$ and $f$.  We show that  when this condition holds it is possible
  to  achieve  the  degenerate  counterpart  of  conventional  STIRAP,
  whereby  adiabatic passage  produces complete  population  transfer. 
  Indeed, the system is equivalent to a set of independent three-state
  systems, in each of which  a STIRAP procedure can be implemented. We
  describe  a scheme  of  unitary transformations  that produces  this
  result.   We  also  examine  the cases  when  this  degeneracy
  constraint does not hold, and show what can be accomplished in those
  cases.  For example, for angular momentum states when the degeneracy
  of the $g$ and $f$ levels is less than that of the $e$ level we show
  how  a special  choice for  the pulse  polarizations and  phases can
  produce complete removal of population from the $g$ set.  Our scheme
  can be a  powerful tool for coherent control  in degenerate systems,
  because of its robustness when selective addressing of the states is
  not required or impossible.  We illustrate the analysis with several
  analytically solvable examples,  in which the degeneracies originate
  from  angular   momentum  orientation,  as   expressed  by  magnetic
  sublevels.
\end{abstract}

\date{\today }


\maketitle

\section{Introduction}

Techniques based  on adiabatic passage provide  very practical methods
for  producing  nearly complete  transfer  of  population between  two
quantum  states  using  crafted  laser pulses  \cite{Vitanov01}.   One
popular  example  of such  coherent  adiabatic excitation,  stimulated
Raman adiabatic passage (STIRAP)  \cite{STIRAP}, provides a simple and
robust technique for transferring population between two nondegenerate
metastable levels,  making use  of two pulses,  termed the  pump pulse
(linking the  initially populated  ground state $\psi_g$  with excited
state $\psi_e$)  and the Stokes pulse (linking  excited state $\psi_e$
with final state $\psi_f$ of  the three-state chain).  When the pulses
are properly  timed (Stokes preceding but overlapping  the pump pulse)
and two-photon resonance is maintained, then via adiabatic passage the
population  is  transferred  from  initial  to  final  state,  without
appreciable population in the excited state at any time.

The operation of STIRAP can be understood by introducing instantaneous
eigenstates  of  the   time-varying  Hamiltonian,  the  time-dependent
adiabatic states with associated time-dependent eigenvalues (adiabatic
energies).   One  (and only  one)  of  these  states, $\Phi_0(t)$,  is
constructed from only  the initial and final state,  with no component
of   the  excited   state.   Because   the  excited   state  generates
fluorescence via  spontaneous emission,  such an adiabatic  state will
exhibit no such  signal; it is termed a {\em  dark} state.  During the
STIRAP  process the state  vector $\Psi(t)$  remains aligned  with the
adiabatic  state  $\Phi_0(t)$,  while  this state,  in  turn,  changes
composition  from  being  aligned  with $\psi_g$  initially  to  being
aligned with $\psi_f$ after the Stokes-pump pulse sequence.

Numerous extensions of the basic three-state STIRAP \cite{STIRAP} have
been  considered \cite{Vitanov01,AAMOP},  including examples  in which
there  occur  magnetic   sublevels  and  associated  degeneracy.   One
possibility is that the atomic energy levels are coupled in such a way
that each one is connected to at most two others.  Population transfer
in  such  multi-state  chains  has  been studied  by  several  authors
\cite{Shore91,   Smith92,   Pillet93,   Weiss94,  Shore95,   Martin95,
  Malinovsky97, Vitanov98, Theuer98}.   In addition to straightforward
population  transfer,  STIRAP  has  been  applied to  the  problem  of
manipulating  and  creating coherent  superpositions  of  two or  more
quantum   states.    Such  superpositions   are   required  for   many
contemporary   applications  including   information   processing  and
communication.   The original  STIRAP process  has, for  example, been
utilized to  create coherent  superpositions in three-  and four-level
systems   \cite{Marte91,  Lawall94,  Weitz94,   Goldner94,  Unanyan98,
  Theuer99, Unanyan99} and to prepare $N$-component maximally coherent
superposition states  \cite{Unanyan01}.  There have  been proposals to
create  $N$-component coherent superpositions  in such  systems, where
the final state  space is degenerate \cite{Kis01, Kis02},  at least in
the rotating  wave picture.  This  idea has been further  developed to
map wave-packets  between vibrational potential  surfaces in molecules
\cite{Kraal02a, Kraal02b}.  Finally, it  has been shown for a specific
degenerate   system,  having   a  single   initial-,   two  degenerate
intermediate-, and three degenerate  final states coupled in the Raman
configuration, that the STIRAP process can be extended to systems with
degenerate intermediate and final levels \cite{Kis03}.

Yet an open question has remained:  what is the most general system of
three degenerate  levels, linked  via Raman process,  for which  it is
possible to transfer all  population from the ground-state manifold of
degenerate states (the  $g$ set) to the final-state  manifold (the $f$
set) while minimizing population in  the excited states (the $e$ set),
without first using optical  pumping to prepare a single nondegenerate
initial state?  We here provide the answer to this question.

We consider $N_g$  degenerate states of the $g$  set, coupled by means
of a  pump-pulse to $N_e$ degenerate  states of the $e$  set, which in
turn are linked by the Stokes  pulse to $N_f$ degenerate states of the
$f$  set.  We  will show  that such  a generalized  STIRAP  process is
almost  always possible  if  the succession  of state-degeneracies  is
nondecreasing, i.e.  $  N_g \leq N_e \leq N_f$.   When such conditions
hold,  then  for arbitrary  couplings  among  states (e.g.   arbitrary
elliptical polarization of  electric dipole radiation between magnetic
sublevels) it is possible to  obtain complete adiabatic passage of all
population from  the states  of the $g$  set into some  combination of
states of the $f$ set.

We  also  examine  the  possibility  of adiabatic  passage  when  this
restriction on degeneracies does not  hold.  We show that in this case
in general only  part of the population can be  transferred to the $f$
set.   We point  out  that, in  special  but important  cases, for  an
appropriate  choice of the  polarizations and  phases of  the coupling
fields, a complete adiabatic population transfer can be obtained.

Another  motivation  of  this   paper  is  the  creation  of  coherent
superposition states  in a degenerate  system. The difficulty  in such
systems  arises from the  limited possibility  of addressing  a single
preselected state: addressing of  a selected state is usually achieved
by exploiting selection rules that  the coupling field should satisfy.
However, if we  have e.g.  two Zeeman multiplets a  light field with a
certain  polarization  will   create  several  couplings  between  the
magnetic sublevels of the multiplets.  Our scheme offers a solution to
this problem: we show that despite of the lack of selective addressing
of  the degenerate  states,  we  have some  control  over the  created
coherent superposition  state in  the $f$ set.   As we point  out, and
illustrate with specific examples, the level of control depends on the
system under consideration.

Our scheme is based on using a Morris-Shore (MS) transformation of the
Stokes  couplings  or  the   pump  couplings,  thereby  reducing  this
particular  (generally  complicated)  linkage  to a  set  of  unlinked
two-state  systems and  dark  states \cite{Morris83,5ss}.   Underlying
this technique is  the fact that, as Morris  and Shore \cite{Morris83}
have  shown, any  system of  linkages in  which there  occur  only two
detunings  (i.e. the  system  has two  sets  of degenerate  sublevels,
termed here $a$  and $b$, forming sets of  dimension $N_a$ and $N_b$),
can be transformed, via suitable  redefinition of basis states, to one
involving  a  set  of  $N_{<}$ independent  two-state  systems,  where
$N_{<}={\rm min}\{N_a, N_b\}$, together with a set of uncoupled states
that are unconnected to other states by the given couplings (one-state
systems).  If  such an uncoupled state  has no component  from the $e$
set  we term  it a  {\em dark}  state.  We  here extend  that  work to
produce sets of unlinked three-state systems.

The paper  is organized as follows:  In the next section  we present a
general model for degenerate, three-level systems and discuss its main
properties.  In  Sec.~\ref{sec:mstrafo} we derive  a general condition
for     complete      STIRAP-like     population     transfer.      In
Sec.~\ref{sec:stokes-mstrafo} we  derive analytic expressions  for the
dark and bright states for  important special choices of degeneracies. 
Then,  in  Sec.~\ref{sec:adiab-tevol},  we  determine  the  conditions
needed  for adiabatic  evolution.  We  demonstrate our  method through
some  specific  examples   in  Sec.~\ref{sec:examples}.   Finally,  in
Sec.~\ref{sec:summary}, we summarize our results.

\section{The Degenerate-Sublevel Model} \label{sec:model}

\subsection{The Hamiltonian}

As is customary when dealing  with STIRAP or other three-level chains,
we  introduce  an  expansion   of  the  state  vector  $\Psi(t)$  that
incorporates  explicit phases  taken from  carrier frequencies  of the
pump and  Stokes pulses, $\omega_p$ and  $\omega_S$, respectively.  In
this  rotating-wave  picture,  and   with  the  customary  neglect  of
counter-rotating   terms   [i.e.    time   variations   $(\omega_i   +
\omega_j)t$] the  rotating-wave approximation (RWA)  Hamiltonian takes
the block-matrix form
\begin{equation}\label{ham}
{\bm H}(t)=\left[\begin{array}{ccc}
        {\bm 0}&p(t){\bm P}&{\bm 0}\\
        p(t){\bm P}^\dagger&\hbar{\bm \Delta}&s(t){\bm S}\\
        {\bm 0}&s(t){\bm S}^\dagger&{\bm 0}
\end{array}\right]\,,
\end{equation}
for use with the Schr\"odinger equation
\begin{equation}
      i\hbar\frac{d}{dt} {\bm C}(t)={\bm H}(t){\bm C}(t)\,.
\end{equation}
Here the zeros ${\bm 0}$ denote null square or rectangular matrices of
appropriate dimensions.   The zero matrix  in the bottom  right corner
indicates  that   the  system  is  supposed   to  maintain  two-photon
resonance.   All time dependence  occurs in  the two  pulse amplitudes
$p(t)$ and $s(t)$, each with  unit maximum value.  The $N_e\times N_e$
diagonal matrix  $\hbar\bm\Delta$ describes  the detuning of  the pump
carrier frequency  from the  Bohr frequency of  the $g-e$  transition. 
The  $N_g\times N_e$  matrix $2  p(t){\bm P}/\hbar$  consists  of Rabi
frequencies associated  with the transitions  between the $g$  and $e$
sets, $\hbar\Omega_{ij}(t)=2p(t)P_{ij}$.  The elements of the constant
matrix ${\bm P}$ read
\begin{equation}
      P_{ij}=\frac{1}{2}{\cal E}^{(p)}\mu_{ij}\,,\qquad
      \left\{\begin{array}{l}
        i=1\hdots N_g \\
        j=1\hdots N_e
      \end{array}\right.\,,
\end{equation}
where  ${\cal  E}^{(p)}$  is  the  peak amplitude  of  the  pump-pulse
electric field and $\mu_{ij}$ is the dipole-transition moment.

Similarly, the $N_e\times N_f$ matrix $2s(t){\bm S}/\hbar$ consists of
Rabi frequencies  associated with the transitions between  the $e$ and
$f$ sets of states.  The elements of the constant matrix $\bm S$ are
\begin{equation}
      S_{ij}=\frac{1}{2}{\cal E}^{(S)}\mu_{ij}\,,\qquad
      \left\{\begin{array}{l}
        i=1\hdots N_e \\
        j=1\hdots N_f
      \end{array}\right.\,,
\end{equation}
where ${\cal  E}^{(S)}$ is the  peak amplitude of the  Stokes electric
field.

The structure of the RWA  Hamiltonian of Eq.~(\ref{ham}) is similar to
that  of  the conventional  three-state  STIRAP,  in  having all  time
dependence confined  to two pulses  $p(t)$ and $s(t)$, but  instead of
single ground, excited, and  final states we have degenerate manifolds
of sublevels,  and hence  we have matrices  $ p(t){\bm  P}$, $s(t){\bm
  S}$, and ${\bm \Delta}$  where conventional STIRAP would have scalar
elements. To illustrate  these Fig.  1 shows the  linkage patterns for
the angular momentum sequence $J = 2 \leftrightarrow 3 \leftrightarrow
4$.   To simplify  the drawings  we  show the  energies of  successive
manifolds as increasing, such as would occur with a ladder scheme; the
connections are the same as  with the usual lambda couplings, in which
the final sublevels have energies below the excited state.

\begin{figure}
    \includegraphics[width=\textwidth/2-1cm]{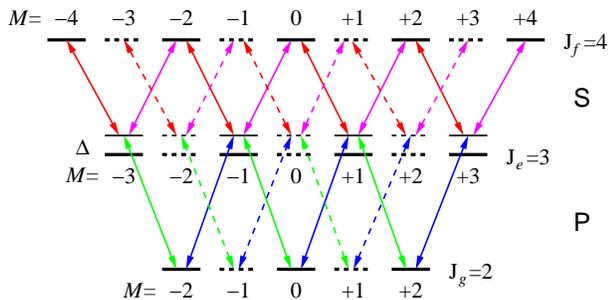} 
    \caption{(Color Online) An example for the degenerate STIRAP scheme:
      we have  three Zeeman multiplets with $J_g=2$,  $J_e=3$, $J_f=4$.
      The couplings  are those of $\sigma_{\pm}$  polarized pulses. The
      pump and Stokes pulses are  detuned from exact resonance with the
      excited-state by $\Delta$, but they maintain two-photon resonance
      between  states  $g$ and  $f$.   The  system  separates into  two
      independent systems, indicated by solid and dashed lines.  }
\label{fig:234-scheme}
\end{figure}

Although we  discuss situations in which the  coupling matrices result
from  magnetic-sublevel degeneracy,  all  of our  results apply  quite
generally, for any mathematical form of the dipole-moment matrices and
consequently  for any  arbitrary  structure of  the constant  matrices
${\bm S}$ and ${\bm P}$.

\subsection{Dark states} \label{sec:dark-states}

There exist  $N=N_g +  N_e +  N_f$ basis states  for this  system, and
hence $N$  adiabatic states $\advecb_n(t)$.  We  can immediately apply
the  MS transformation \cite{Morris83},  at each  instant of  time, by
placing the $g$  and $f$ sets of states together into  the MS $a$ set,
and taking the $e$ set to be the MS $b$ set.  If the $a$ set is larger
than the $b$  set, there will be  $N_u = N_a - N_b$  uncoupled states.
None of these have any component from the $e$ set, and so they are all
dark states.   The number of dark  states is thus  $N_D = N_g +  N_f -
N_e$.   In  the conventional  nondegenerate  STIRAP \cite{AAMOP},  for
which $N  = 3$,  the MS  transformation gives one  dark state  and one
bright state; for the tripod system,  for which $N = 4$, there are two
dark states \cite{Unanyan98,Theuer99}.  In the angular-momentum system
of Fig.   \ref{fig:234-scheme} there are $N_D  = 5 +  9 - 7 =  7$ dark
states.

For  conventional nondegenerate  STIRAP  the composition  of the  dark
state  changes with  time, because  the coupling  matrices and  the MS
transformation change with time.  However, it is possible to associate
the (single) dark state  initially with the nondegenerate ground state
by  applying the  pulses in  the counterintuitive  order,  i.e. Stokes
pulse preceding pump pulse.  When there is degeneracy, it is necessary
to establish that  the entire population of any  pure initial state in
the $g$ set is projected into the set of dark states and no population
is left in bright states.  This  completeness of the dark states is at
the heart  of our question  concerning the possibility of  STIRAP with
degeneracy.

\section{General condition for  complete population transfer}
\label{sec:mstrafo}

One of our basic questions is whether, for a given linkage pattern, it
is possible to empty completely  the $g$ set for any arbitrary initial
state, once we have fixed the pump and Stokes pulses.

It is easy to see that one necessary condition for complete removal of
population from the  ground manifold is that there  should not be more
sublevels in this  manifold than there are in the  excited state, i.e.
we require $N_g \leq N_e$

To prove this assertion  we employ a MS transformation \cite{Morris83}
on the pump transitions that  connect ground and excited states.  This
transformation introduces a  new set of basis states  in each of these
manifolds, such that each sublevel from the $g$ set couples to at most
one sublevel  from the  $e$ set.  Were  there are no  Stokes couplings
between $e$ and  $f$ states, the dynamics could be  described as a set
of  independent two-state  systems, together  with some  single states
(uncoupled states) that are not affected by the pump radiation.  Given
such a revision of  the basis states, it is easy to  see that if there
are more ground states than excited states, $N_g > N_e$, then the dark
states  will be  composed of  $g$-states and  some population  will be
trapped  there.   This  will   remain  unaffected  by  the  radiation;
population cannot  be removed from them using  this particular linkage
pattern.

\begin{figure}
      \includegraphics[width=5cm]{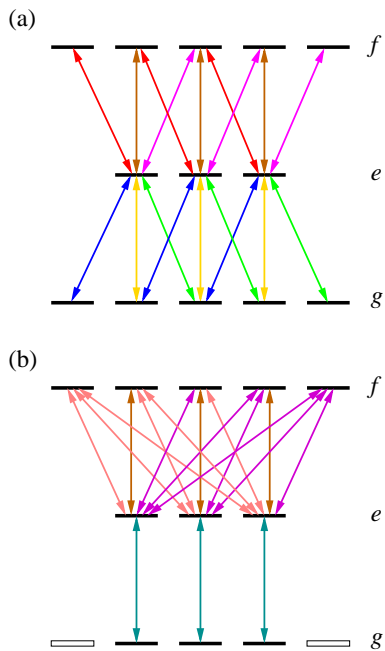} 
\caption{(Color Online)
   This  sketch shows  that when  the number  of ground-state-sublevels
   exceeds the number of  excited-state-sublevels then it is impossible
   to transfer all the population from the ground-state manifold to the
   final-state  manifold.    Frame~(a)  shows  the   original  coupling
   configuration.    Frame~(b)   shows  the   couplings   after  a   MS
   transformation  on  the  $g-e$  transition.   The  empty  rectangles
   represent uncoupled  states.  This transformation  is independent of
   time,  because all  elements of  the  coupling share  a common  time
   dependence, $p(t)$.  The presence  of uncoupled sublevels in the $g$
   set prevents  removal of population  from these states; hence  it is
   not possible to  remove all population from all  of the ground-state
   sublevels. }
\label{fig:212scheme}
\end{figure}

Figure \ref{fig:212scheme} illustrates  this accounting procedure. The
top frame (a)  shows a general coupling scheme for  the sequence $J= 2
\leftrightarrow  1 \leftrightarrow  2$. The  MS transformation  on the
$g-e$  pump transition produces  the description  shown in  the bottom
frame (b).  In  the $g$ set, with this  transformed basis, there occur
two  sublevels  that have  no  connection  with  any excited  states.
Population cannot be  removed from these as long  as the couplings are
those shown in the top frame.

It is easy to see that, had  there been more sublevels in the $e$ set,
such that $N_g \le N_e$, then every one of the transformed states from
the $g$  set would  be linked to  some excited state,  with consequent
possibility  for population  removal.   There will  also be  uncoupled
states in the $e$ manifold but  they are unpopulated and do not affect
the population transfer.

The  introduction of  MS  basis states  in  this way  makes the  $g-e$
linkage pattern quite simple, but by introducing a new basis the $e-f$
couplings  become   more  complicated:  generally  there   will  be  a
connection between each transformed $e$ state and each (untransformed)
$f$ state, as indicated in frame (b).

Next  we consider  the  $e-f$  coupling. We  can  repeat the  previous
argument    for   the   $g-e$    coupling   with    the   replacements
$g\leftrightarrow e$  and $e\leftrightarrow  f$.  We obtain,  that the
Stokes field  MS transformation yields  $N_{<}=\mbox{min}\{N_e, N_f\}$
independent  two   level  systems  for  the   $e-f$  transition,  plus
$|N_e-N_f|$ uncoupled states in the larger  one out of the $e$ and $f$
sets. It is easy to see that, had there been more sublevels in the $f$
set, such that $N_e \le N_f$, then every one of the transformed states
from the $e$ set would be  linked to some final state, with consequent
possibility for  population removal. Combinig the arguments  of the MS
tarnsormations for  the $g-e$ and  $e-f$ couplings, we obtain  that in
general,  if  a  non--descending  sequence of  state--degeneracies  is
fulfilled
\begin{equation}\label{cond}
      N_g\leq N_e \leq N_f\,,
\end{equation}
then a  complete STIRAP-like population  transfer from the $g$  set to
the $f$ set  is possible.  We emphasize that in  this case the success
of  the full  transfer  is independent  of  the initial  state of  the
system: it can be any pure state or a mixed state as well.

A particularly important special  case of degeneracy occurs when there
are  dark  states  but  they  are  insufficient  to  produce  complete
population transfer.   This occurs when $N_g+N_f>N_e$,  but $N_f<N_e$.
For example,  in the linkage of  $J=1\leftrightarrow 2 \leftrightarrow
1$ there is 1 dark state.  Figure~\ref{fig:121scheme} illustrates this
situation.

\section{The Stokes-field MS transformation} \label{sec:stokes-mstrafo}

In this  section we  determine the dark  states of the  Hamiltonian of
Eq.~(\ref{ham}); these are the  adiabatic states that will be utilized
for the  desired adiabatic population transfer.  In  order to simplify
the structure of the Hamiltonian, we perform a MS transformation; here
we take that to be on the $e-f$ couplings (those of the Stokes field).
In our  case the time-independent  transformation matrix ${\bm  U}$ is
defined as
\begin{equation}\label{Udef}
{\bm U}=\left[
      \begin{array}{ccc}
        {\bm I}&{\bm 0}&{\bm 0}\\
        {\bm 0}&{\bm B}&{\bm 0}\\
        {\bm 0}&{\bm 0}&{\bm A}
      \end{array}\right]\,.
\end{equation}
In the top-left  corner there is a unit matrix  ${\bm I}$ of dimension
$N_g\times N_g$.   This leaves the  $g$ set of states  unaltered.  The
$N_f\times N_f$ unitary matrix $\bm A$ transforms the sublevels in the
final-state manifold.   Similarly, the $N_e\times  N_e$ unitary matrix
$\bm B$  transforms the sublevels in the  excited-state manifold.  The
constant matrices $\bm A$ and $\bm B$ are defined \cite{Morris83} such
that by  transforming the Hamiltonian Eq.~(\ref{ham})  with the matrix
$\bm U$ through the relation
\begin{equation}\label{trafo}
{\bm U}{\bm H}(t){\bm U}^\dagger = \left[\begin{array}{ccc}
        {\bm 0}&p(t)\widetilde{\bm P}&{\bm 0}\\
        p(t)\widetilde{\bm P}^\dagger&\hbar{\bm\Delta}
        &s(t)\widetilde{\bm S}\\
        {\bm 0}&s(t)\widetilde{\bm S}^\dagger
        \end{array}\right]\,
\end{equation}
we  obtain a  transformed pump-field  coupling  matrix $\widetilde{\bm
  P}={\bm  P}{\bm  B}^\dagger$,   and  a  quasi-diagonal  Stokes-field
coupling  matrix  $\widetilde{\bm  S}=\bm  B\bm S\bm  A^\dagger$.   By
quasi-diagonal we mean that the structure of the matrix is
\begin{eqnarray}\label{Str}
      \widetilde{\bm S}  =
      \begin{cases} \left[\begin{array}{cc}
            \widetilde{\bm \Sigma} &
            {\bm 0} \end{array}\right] &
             \mbox{ if }N_f>N_e\,,
             \\[6pt]
      \widetilde{\bm \Sigma} &
            \,\,\mbox{if } N_f=N_e\,,
            \\[2pt]
      \left[\begin{array}{c}
          \widetilde{\bm \Sigma} \\ {\bm 0}
          \end{array}\right] &
      \mbox{ if } N_f<N_e\,,
      \end{cases}
\end{eqnarray}
where  $\widetilde{\bm  \Sigma}$  is  a square  diagonal  matrix  with
dimension  $N_<={\rm  min}(N_e, N_f)$.   The  moduli  of the  diagonal
elements are  given by the  square-roots of the common  eigenvalues of
the Hermitian  matrices $\bm  S\bm S^{\dag}$ (of  dimension $N_e\times
N_e$)  and $\bm S^{\dag}\bm  S$ (of  dimension $N_f\times  N_f$).  The
phases of  the diagonal elements  are obtained by  evaluating directly
the matrix  product $\bm B\bm  S\bm A^\dagger$.  Some of  the diagonal
elements of $\widetilde{\bm \Sigma}$  might be zero, meaning that some
$e-f$  couplings vanish  in  the MS  basis.   We here  assume that  in
general all diagonal elements of $\widetilde{\bm\Sigma}$ are non-zero,
i.e. it is nonsingular.  We treat in Appendix \ref{sec:sing-sigma} the
case when this matrix is singular.

In the  following subsections we consider the  three important special
cases of degeneracies  and derive the adiabatic states  of the coupled
degenerate systems.

\subsection{The case $ N_g\leq N_e\leq N_f$}\label{sec:piramid1}

\begin{figure}
      \includegraphics[width=5cm]{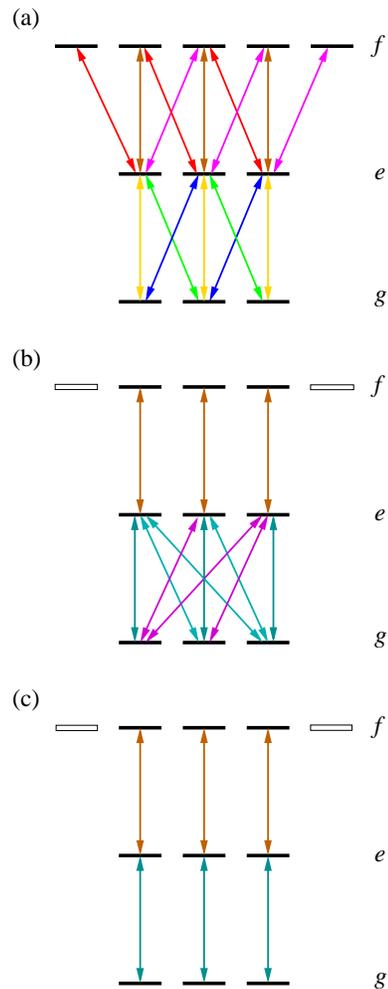} 
      \caption{(Color Online)
        The  three stages  of  the transformations.   (a) The  original
        coupling  scheme.   (b)  The  result  of  the  Stokes-field  MS
        transformation,  converting the couplings  between $e$  and $f$
        sets  into independent  one-  and two-state  systems.  (c)  The
        result of redefining the states in  the $g$, $e$, and $f$ sets.
      }
\label{fig:112scheme}
\end{figure}

We first  consider the  case when the  MS transformation on  the $e-f$
transition  results  in $N_f-N_e>0$  decoupled  sublevels  in the  $f$
manifold.   The coupling  matrix  $\widetilde{\bm S}$  takes the  form
given in the  first row of Eq.~(\ref{Str}), and  hence the Hamiltonian
in the MS basis reads
\begin{equation}\label{ham-ms}
    \widetilde{{\bm H}}(t)=\left[\begin{array}{cccc}
        {\bm 0}&p(t)\widetilde{{\bm P}}&{\bm 0}&{\bm 0}\\
        p(t)\widetilde{{\bm P}}^\dagger&\hbar{\bm
          \Delta}&s(t)\widetilde{\bm \Sigma}
        &{\bm 0}\\
        {\bm 0}&s(t)\widetilde{\bm \Sigma}^\dagger&{\bm 0}&{\bm 0}\\
        {\bm 0}&{\bm 0}&{\bm 0}&{\bm 0}
      \end{array}\right]\,.
\end{equation}
As with the original RWA  Hamiltonian, the only time dependence enters
through the pulses $p(t)$ and $s(t)$.

We can  treat the  system in  the same way  when $N_f=N_e$.   Then the
coupling  matrix $\widetilde{\bm  S}$ is  given by  the second  row of
Eq.~(\ref{Str}), and  we have to omit  all zero rows  and columns from
the Hamiltonian of Eq.~(\ref{ham-ms}).  In either cases the sub-matrix
$\widetilde{{\bm P}}$ has dimensions $N_g\times N_e$, while the square
matrices  $\widetilde{\bm  \Sigma}$  and $\bm\Delta$  have  dimensions
$N_e\times N_e$.

To  find  the  adiabatic  eigenvectors  $\widetilde{\advecb}_k(t)$  of
$\widetilde{{\bm H}}(t)$ we take their elements to have the form
\begin{equation}
    \label{vkparam}
    \widetilde{\advecb}_k(t) = \left[\begin{array}{c}
        {\bm x}_k \t  \\
        \widetilde{\bm y}_k \t \\
        \widetilde{\bm z}_k \t \\
        \widetilde{\bm z}'_k \t
      \end{array}\right]\quad
    \begin{array}{c}
      {g} \\ {e} \\ {f} \\ {f'}
    \end{array}
\end{equation}
where $f'$  denotes the subspace of  uncoupled states in the  $f$ set.
Because these are unlinked to the  $e$ set they meet the definition of
dark  states.  Their  population, if  initially present,  is preserved
throughout  the time  evolution.  When  $N_f=N_e$ we  simply  omit the
fourth row  from this vector (the  $f'$ states), i.e.  we  do not have
$\widetilde{\bm z}'_k$.   In Eq.~(\ref{vkparam}) there is  no tilde on
the $x$ components  because, unlike the $y$ and  $z$ components, these
do not transform in the Stokes  field MS transformation.  In Sec. IV B
and C the $x$ components  undergo a MS transformation, as is indicated
there by a tilde.

The eigenvectors satisfy the eigenvalue equation
\begin{equation}\label{eigeneqdef}
    \widetilde{\bm {H}}(t)\widetilde{\advecb}_k(t)=\eigenv_k(t)
\widetilde{\advecb}_k(t)\,.
\end{equation}
By  substituting   the  Hamiltonian  of   Eq.~(\ref{ham-ms})  and  the
parameterization  (\ref{vkparam})   of  the  eigenvectors   into  this
equation we  obtain four  sets of coupled  linear equations  for ${\bm
  x}_k$,    $\widetilde{\bm   y}_k$,   $\widetilde{\bm    z}_k$,   and
$\widetilde{\bm z}'_k$.   The solution of these  equations provide the
dark  and bright  eigenvectors  $\widetilde{\advecb}_k(t)$ defined  by
Eq.~(\ref{eigeneqdef}).

Let us assume  that there exists an eigenvalue  zero, $\eigenv_0=0$\,. 
This is always possible to ensure, by suitable choice of the phases of
the rotating wave  approximation and the zero-point of  energy.  If we
can find a solution  of the eigenvalue-equation (\ref{eigeneqdef}) for
this case, then our assumption $\eigenv_0=0$ holds, since the solution
of the linear equations is  unique.  After some algebra one can obtain
$N_g$ different  vectors $\widetilde{\advecb}^{(l)}_0 \t\,\,,l=1\ldots
N_g$, that are linearly independent  of each other, and can make these
orthonormal
\begin{equation}\label{darkstates1}
   \widetilde{\advecb}_0^{(l)}(t)= {1\over {\cal N}_0^{(l)}(t)}\left[
     \begin{array}{c}
       s(t){\bm x}_0^{(l)} \\  {\bm 0} \\
       -p(t)\widetilde{\bm \Sigma}^{-1}\widetilde{{\bm
           P}}^\dagger{\bm x}_0^{(l)} \\
       {\bm 0}
     \end{array}\right]\,,
\end{equation}
where  ${\cal  N}_0^{(l)} (t)$  is  a  (time dependent)  normalization
factor.  Here we have assumed that the matrix $\widetilde{\bm \Sigma}$
is   nonsingular.    We   will   discuss   separately,   in   Appendix
\ref{sec:sing-sigma}, the  situation when $\widetilde{\bm  \Sigma}$ is
singular. Since the $y$ component  of these vectors is zero,   they
have no component in the $e$  set; they correspond to dark states.  To
make  the dark eigenvectors  of Eq.~(\ref{darkstates1})  orthogonal we
require that
\begin{equation}\label{ortho}
   s(t)^2 \langle{\bm x}^{(k)\,T}_0|{\bm x}^{(l)}_0\rangle+
   p(t)^2\langle{\bm x}^{(k)\,T}_0|
   \widetilde{{\bm P}}\widetilde{\bm \Sigma}^{-1\dagger}
   \widetilde{\bm \Sigma}^{-1}\widetilde{{\bm P}}^\dagger
   |{\bm x}^{(l)}_0\rangle=0\,,
\end{equation}
for $1  \leq k  < l  \leq N_g$.  The  time-dependence of  the envelope
functions  $s(t)$ and $p(t)$  is arbitrary,  and therefore  we require
that the two terms on the left-hand-side (lhs) of Eq.~(\ref{ortho}) be
identically  zero.  The  eigenvectors  of a  Hermitian  matrix can  be
chosen so  that they are  orthogonal to each-other, and  therefore the
first term on the lhs  of Eq.~(\ref{ortho}) is automatically zero.  It
follows that the vectors ${\bm x}^{(l)}_0$ are the eigenvectors of the
Hermitian matrix
\begin{equation}\label{metric}
   {\bm M} = {\bm P}({\bm S}{\bm S}^{\dagger})^{-1}{\bm P}^{\dagger}\equiv
   \widetilde{{\bm  P}}\widetilde{\bm
     \Sigma}^{-1\dagger}\widetilde{\bm  \Sigma}^{-1}
   \widetilde{{\bm  P}}^\dagger\,.
\end{equation}

There is another set of dark eigenvectors for $N_f>N_e$.  These follow
from the discussion after Eq.~(\ref{vkparam}) and are given by
\begin{equation}
   \widetilde{\advecb}_0^{(l)}=\left[
     \begin{array}{c}
        {\bm 0} \\  {\bm 0} \\ {\bm 0} \\ {\bm z}^{\prime\, (l)}
     \end{array}\right]\,,\quad l=N_g+1, \ldots N_f-N_e+N_g\,,
\end{equation}
where ${\bm z}^{\prime\, (l)}$  are constant orthonormal unit vectors. 
These   dark  eigenvectors   are  clearly   orthogonal  to   those  of
Eq.~(\ref{darkstates1}).

We show  in Appendix~\ref{sec:linearized-couplings} that  the coupling
sequence  $g\leftrightarrow  e\leftrightarrow f$  can  be rendered  to
independent three-state chains by a suitable hoice of the basis states
in the $g$, $e$, and $f$ sets.  Figure~\ref{fig:112scheme} illustrates
the sequence of transformations that  leads to the construction of the
dark-state eigenvectors Eq.~(\ref{darkstates1}).   Frame (a) shows the
original system,  with some couplings.  Frame (b)  depicts the results
of  the Stokes-field  MS transformation  of the  $e$ and  $f$  states. 
Frame (c)  shows the result of  the redefinition of the  $g$, $e$, and
$f$ sets  of states according to Eq.~(\ref{sets}),  with the resulting
set of independent chains.

The matrix  of Eq.~(\ref{metric}) may  have zero eigenvalues as  well.
If so,  the corresponding  eigenvectors ${\bm x}_0^{(k)}$  satisfy the
equation
\begin{equation}\label{coupcond}
   {\bm P}^{\dagger}{\bm x}_0^{(k)}={\bm 0}\,,
\end{equation}
since  we have  assumed that  the matrix  $\widetilde{\bm  \Sigma}$ is
nonsingular.   Note that  here ${\bm  P}^{\dag}$ is  expressed  in the
original   atomic  basis.   The   $i$th  row   of  the   matrix  ${\bm
   P}^{\dagger}$ describes the coupling  between state $i$ from the $e$
set and the sublevels of the  $g$ set. The rows of the coupling matrix
can be considered  as vectors that span a subspace  of states from the
$g$ set.   The dimension  of this subspace  is the number  of linearly
independent rows of ${\bm P}^{\dagger}$, say $N_P$.  Obviously we have
$N_P\leq  N_e$  and $N_P\leq  N_g$.   Therefore,  there are  $N_g-N_P$
different,   nontrivial  solutions  of   Eq.~(\ref{coupcond}).   These
nontrivial solutions  provide states that  are unaffected by  the pump
field.

If $N_g=N_P\leq N_e$ then such  an uncoupled state does not exist, and
the  vectors $\{{\bm  x}_0^{(k)}\}$,  $k=1\ldots N_g$  span the  total
$g$-set   manifold.    Therefore   by  choosing   a   counterintuitive
pulse-sequence for the  pump and Stokes pulses, we  can cause complete
transfer of  population from the  $g$ set to  the $f$ set by  means of
independent  STIRAP  processes.    For  such  population  transfer  to
succeed,  the   conditions  of  the  adiabatic   evolution  should  be
fulfilled,  as  we will  discuss  in Sec.~\ref{sec:adiab-tevol}.   The
success  of such  population transfer  is independent  of  the initial
state  of  the system.   It  can be  any  single  state, an  arbitrary
coherent  superposition  of  states   or  even  a  mixed  state,  {see
   Sec.~\ref{sec:adiab-tevol}}.

If $N_P<N_g$ then  some $g$-set sublevels are decoupled  from the pump
field,  hence  in  general it  is  then  impossible  to move  all  the
population from the $g$ set.  Part of it is trapped in dark states.

The   other   $2N_e$  adiabatic   eigenvectors   belong  to   non-zero
eigenvalues. They can be obtained in the form
\begin{equation}
   \widetilde{\advecb}_k \t= {1\over {\cal N}_k(t)}\left[\begin{array}{c}
       p(t)\widetilde{{\bm P}} \widetilde{\bm y}_k \t\\
       \eigenv_k \t \widetilde{\bm y}_k \t\\
       s(t)\widetilde{\bm \Sigma}^{\dagger} \widetilde{\bm y}_k \t\\
       {\bm 0}
     \end{array}\right]\,,\qquad k=1\ldots 2N_e
   \label{eq:vk}
\end{equation}
where ${\cal  N}_k(t)$ is a normalization factor  and $\widetilde{\bm y}_k
\t$ satisfies the eigenvalue equation
\begin{equation}
   \left[ p(t)^2\widetilde{{\bm P}}^\dagger \widetilde{{\bm P}} + v(t)^2
     \widetilde{\bm \Sigma}\widetilde{\bm \Sigma}^{\dagger}\right]
   \widetilde{\bm y}_k \t =
   \eigenv_k\t [\eigenv_k\t-\hbar\Delta] \widetilde{\bm y}_k \t\,.
   \label{eq:eigen}
\end{equation}
Because  they contain  component states  from the  $e$ set,  these are
bright  states.  Although  for  population transfer  we  use the  dark
states of  Eq.~(\ref{darkstates1}), we need the bright  states to find
the adiabaticity conditions; see Sec.~\ref{sec:adiab-tevol}.

In  summary: in  this  subsection  we have  shown  that when  $N_g\leq
N_e\leq N_f$,  under very  general conditions the  complete population
from the $g$ set can be transferred to the $f$ set of states.  Once we
have  fixed  the  pulse-shapes,  polarizations  and  phases,  complete
transfer can be obtained for  any arbitrary initial state from the $g$
set.  The eigenvectors $\advecb_k\t$, $k=0\ldots 2N_e$ in the original
bare atomic basis can be obtained as
\begin{equation}\label{advecb}
   \advecb_k\t = \frac{1}{{\cal N}_k\t}{\bm U}^\dagger\left[ \begin{array}{c}
       {\bm x}_k \t \\
       \widetilde{\bm y}_k\t \\
       \widetilde{\bm z}_k\t \\
       {\bm 0} \\
     \end{array} \right] =\frac{1}{{\cal N}_k\t}
   \left[ \begin{array}{c}
       {\bm x}_k \t\\
       {\bm B}^\dagger \widetilde{\bm y}_k\t\\
       {\bm A}^\dagger \left[ \begin{array}{c}
           \widetilde{\bm z}_k\t \\
           {\bm 0}
         \end{array} \right]
     \end{array} \right]\,.
\end{equation}
Moreover,  with  this method  it  is  possible  not only  to  transfer
populations, but  to create superposition  states in the $f$  set.  We
will consider this possibility in Sec.~\ref{sec:examples}.

\subsection{The case $N_g>N_e>N_f$}\label{sec:piramid2}

According  to  the  considerations   presented  in  the  beginning  of
Sec.~\ref{sec:mstrafo}, we cannot expect  that all the population from
the $g$ set can be removed when $N_g>N_e>N_f$.  However, a part of the
population  can  be removed  and  with  this  we can  create  coherent
superposition states in  the $f$ set.  In order to  find the dark- and
bright  states  of  the system  we  proceed  in  the  same way  as  in
Sec.~\ref{sec:piramid1}, but now  with the MS transformation involving
the pump transition
\begin{equation}\label{Udef2}
   {\bm U}=\left[
     \begin{array}{ccc}
       {\bm B}&{\bm 0}&{\bm 0}\\
       {\bm 0}&{\bm A}&{\bm 0}\\
       {\bm 0}&{\bm 0}&{\bm I}
     \end{array}\right]\,.
\end{equation}
We look for the eigenvectors of the transformed Hamiltonian in the form
\begin{equation}
   \label{vkparam2}
   \widetilde{\advecb}_k \t = \left[\begin{array}{c}
       \widetilde{\bm x}_k \t \\
       \widetilde{\bm x}^{\prime}_k \t \\
       \widetilde{\bm y}_k \t \\
       {\bm z}_k \t
     \end{array}\right]\,.\quad
    \begin{array}{c}
      {g} \\ {g'} \\ {e} \\ {f}
    \end{array}
\end{equation}
The vectors $\widetilde{\bm  x}^{\prime}_k \t$ describe the population
in those states of the $g$ set that are decoupled from the pump field.
There are $N_g-N_e$ dark states in  the $g$ manifold, and these can be
written in the form
\begin{equation}
\label{vkparam2b}
\widetilde{\advecb}_0^{(l)} \t= \left[\begin{array}{c}
        {\bm 0}\\
        \widetilde{\bm x}^{\prime (l)}_0 \t \\
        {\bm 0}\\
        {\bm 0}
        \end{array}\right]\,,
\end{equation}
where the vectors $\{\widetilde{\bm  x}^{\prime (l)}_0 \t\} $ form
an orthonormal  set.
The population cannot  be removed from these states.   The rest of the
dark    states    are    obtained    in   the    manner    used    for
Eq.~(\ref{darkstates1}). They can be written as
\begin{equation}
      \widetilde{\advecb}^{(k)}_0 \t = {1\over {\cal N}^{(k)}_0 \t} \left[
        \begin{array}{c}
        s(t)\widetilde{\bm \Pi}^{\dag -1}\widetilde{{\bm S}}{\bm z}^{(k)}_0 \\
        {\bm 0} \\
        {\bm 0} \\
        -p(t){\bm z}^{(k)}_0
        \end{array}\right]\,,
\label{darkstates2}
\end{equation}
where   $\left[\begin{array}{c}    \widetilde{\bm   \Pi}\\   {\bm   0}
   \end{array} \right] =\bm  B\bm P\bm A^\dagger$, with $\widetilde{\bm
   \Pi}$ a  diagonal coupling matrix of dimension  $N_e\times N_e$, and
$\widetilde{{\bm  S}}={\bm A}{\bm S}$.   We require  orthogonality for
the dark  states Eq.~(\ref{darkstates2}).  Hence  the constant vectors
${\bm  z}^{(k)}_0$ are  chosen so  that  they are  eigenstates of  the
Hermitian    matrix    $\widetilde{{\bm    S}}^{\dagger}\widetilde{\bm
   \Pi}^{-1}  \widetilde{\bm \Pi}^{\dagger -1}\widetilde{{\bm  S}}$, in
direct  analogy with the  way the  constant vectors  ${\bm x}^{(k)}_0$
were chosen earlier in Sec.~\ref{sec:piramid1}.

\subsection{The case $N_g,N_f<N_e$}\label{sec:diamond}

\begin{figure}
      \includegraphics[width=5cm]{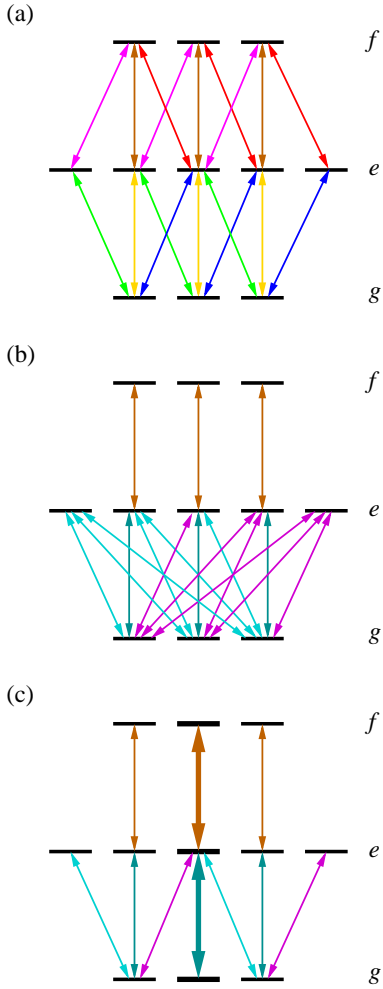} 
\caption{(Color Online) The three frames show the stages of the
   MS transformations  when the sequence of  degenerate states violates
   the  condition  Eq.~(\ref{cond}):  Frame  (a) depicts  the  original
   coupling scheme , here $N_f<N_e$.  Frame (b) shows the result of the
   Stokes field MS transformation, converting the couplings between $e$
   and  $f$ sets  into  independent one-  and  two-state systems.   The
   one-state systems are in the $e$ set.  Frame (c) shows the result of
   the pump field MS transformation followed by the redefinition of the
   sublevels   in   the  $g$,   $e$,   and   $f$   sets  according   to
   Eq.~(\ref{sets}),  leading to one  dark state.   This dark  state is
   associated with  the middle three-state linkage,  indicated by heavy
   lines.  In addition,  the middle $e$ state is coupled  not only to a
   single $g$  state but to the two  others as well.  As  a result, the
   populations in the two spectator $g$ states may disturb the complete
   population transfer from the middle $g$ state, see text.  }
\label{fig:212b}
\end{figure}

Here we consider the situation $N_g,N_f<N_e$.  We will show that under
these conditions  the dark states of  the system can  be identified by
means   of  two   sequential   MS  transformations.    The  first   MS
transformation is performed  among the $e$ and $f$  sets of the Stokes
transition, as  in subsection \ref{sec:piramid1}.   The transformation
matrix is given by Eq.~(\ref{Udef}).  As a result, the coupling matrix
${\bm S}$ of the Hamiltonian (\ref{ham}) takes the quasi-diagonal form
of the  third row of  Eq.~(\ref{Str}).  Therefore, the  Hamiltonian in
the MS basis reads
\begin{equation}\label{Hamb}
   \widetilde{{\bm H}}(t)=\left[\begin{array}{cccc}
       {\bm 0}&p(t)\widetilde{{\bm P}}_a&p(t)\widetilde{{\bm P}}_b&{\bm 0}\\
       p(t)\widetilde{{\bm P}}_a^\dagger&\hbar{\bm \Delta}&{\bm 0}&
       s(t)\widetilde{\bm \Sigma}\\
       p(t)\widetilde{{\bm P}}_b^{\prime\dagger}&{\bm 0}&\hbar{\bm
         \Delta}&{\bm 0}\\
       {\bm 0}&s(t)\widetilde{\bm \Sigma}^\dagger&{\bm 0}&{\bm 0}\\
     \end{array}\right]\,.
\end{equation}
The  diagonal  square matrix  $\widetilde{\bm  \Sigma}$ has  dimension
$N_f\times  N_f$.  It  can be  readily seen  that there  are $N_e-N_f$
states in the  $e$ set that are  not coupled to the $f$  set.  We call
these  {\em uncoupled}  levels, whereas  the other  subset  of coupled
excited-state-sublevels are  called {\em active}.   In the Hamiltonian
of Eq.~(\ref{Hamb})  the pump coupling matrix is  partitioned into two
sub-matrices:   the  matrix   $\widetilde{{\bm  P}}_a$   of  dimension
$N_g\times N_f$ describes couplings between the $g$ set and the active
MS  states of  the  $e$ set.   The  other sub-matrix  $\widetilde{{\bm
     P}}_b$ of  dimension $N_g\times (N_e-N_f)$ is  associated with the
transitions between the states of the $g$ set and the uncoupled states
of the $e$  set.  The result of this  transformation is illustrated in
Fig.~(\ref{fig:212b}b).   As  the  figure  shows, we  cannot  identify
clearly the dark states, because in  general all the states of the $g$
set are coupled to all of the $e$ set.
Therefore, we  perform a second  MS transformation, involving  the $g$
set and just  those states of the $e$ set that  are decoupled from the
$f$ set --  two in the present example.  The  result is illustrated in
Fig.~(\ref{fig:212b}c).  In  this example  there is one  $g$-set state
that couples solely  to an active MS state of the  $e$ set because the
other two have, by means of  the MS transformation, been linked to the
two uncoupled $e$  states.  The population can be  moved from this $g$
state to an  $f$ state.  The middle $e$ state is  coupled to all three
$g$  states.  Consequently,  if  the  two  spectator  $g$  states  are
populated,  they disturb  the  complete population  transfer from  the
middle $g$ state into $f$  states, and the population transfer process
will place population into the $e$ and $f$ states.

In general, the transformation  matrix of the second MS transformation
is defined as
\begin{equation}\label{Updef}
   {\bm U}'=\left[
     \begin{array}{cccc}
       {\bm A}'&{\bm 0}&{\bm 0}&{\bm 0}\\
       {\bm 0}&{\bm I}&{\bm 0}&{\bm 0}\\
       {\bm 0}&{\bm 0}&{\bm B}'&{\bm 0}\\
       {\bm 0}&{\bm 0}&{\bm 0}&{\bm I}
     \end{array}\right]\,.
\end{equation}
The $N_g\times N_g$ unitary matrix  ${\bm A}'$ transforms the $g$ set,
whereas  the  $(N_e-N_f)\times (N_e-N_f)$  unitary  matrix ${\bm  B}'$
transforms the uncoupled states of the $e$ set.  The two unit matrices
${\bm  I}$ are of  dimension $N_f\times  N_f$.  For  $N_e-N_f<N_g$ the
transformation yields
\begin{subequations}\label{pi1}
   \begin{eqnarray}
     {\bm A}'\widetilde{\bm P}_a &=&
     \left[\begin{array}{c}
         \widetilde{\bm P} \\ \widetilde{\bm P}'
       \end{array}\right]\,, \\
     {\bm A}'\widetilde{\bm P}_b {\bm B}^{\prime\dag} &=&
     \left[\begin{array}{c}
         {\bm 0} \\ \widetilde{\bm \Pi}
       \end{array}\right]\,,
   \end{eqnarray}
\end{subequations}
where    the   matrix    $\widetilde{\bm   P}$    is    of   dimension
$[N_g-(N_e-N_f)]\times  N_f$,  $\widetilde{\bm  P}'$ is  of  dimension
$(N_e-N_f)\times N_f$, and $\widetilde{\bm  \Pi}$ is a diagonal matrix
of dimension $(N_e-N_f)\times (N_e-N_f)$. For $N_e-N_f=N_g$ we find
\begin{subequations}\label{pi2}
   \begin{eqnarray}
     {\bm A}'\widetilde{\bm P}_a &=& \widetilde{\bm P}'\,, \\
     {\bm A}'\widetilde{\bm P}_b {\bm B}^{\prime\dag} &=&
     \widetilde{\bm \Pi}\,,
   \end{eqnarray}
\end{subequations}
where the matrix $\widetilde{\bm  P}'$ is of dimension $N_g\times N_f$
and $\widetilde{\bm \Pi}$ is of  dimension $N_g\times N_g$.  We do not
have $\widetilde{\bm P}$ in  this case.  Finally, for $N_e-N_f>N_g$ we
get
\begin{subequations}\label{pi3}
   \begin{eqnarray}
     {\bm A}'\widetilde{\bm P}_a &=& \widetilde{\bm P}'\,, \\
     {\bm A}'\widetilde{\bm P}_b {\bm B}^{\prime\dag} &=&
     \left[\begin{array}{cc}
         \widetilde{\bm \Pi}&{\bm 0}
       \end{array}\right]\,,
   \end{eqnarray}
\end{subequations}
where the matrix $\widetilde{\bm  P}'$ is of dimension $N_g\times N_f$
and $\widetilde{\bm  \Pi}$ is of  dimension $N_g\times N_g$.   Just as
with  the   conditions  $N_e-N_f=N_g$,  we   do  not  have   a  matrix
$\widetilde{\bm P}$ in the present case either.

In   general,  none   of   the  diagonal   elements   of  the   matrix
$\widetilde{\bm  \Pi}$ are  zero.  Therefore,  when  $N_e-N_f\geq N_g$
there are no  states in the $g$ set that are  coupled solely to active
MS  states in  the $e$  set.  It  follows that  no dark  state  can be
identified in  the system and hence a  STIRAP-like population transfer
is impossible.

In  special (but  important) cases  it  may occur  that some  diagonal
elements of $\widetilde{\bm \Pi}$ vanish.  Then the system has such MS
$g$-set states  that are coupled only  to active MS states  of the $e$
set,  hence  the  system has  dark  states  and  a STIRAP  process  is
possible. We will reconsider this case later in this subsection.

In  all three  cases $N_e-N_f>N_g$,  $N_e-N_f=N_g$,  and $N_e-N_f<N_g$
some initial population of the $g$  set cannot be included in the dark
states  of  the  system  in  the general  case  of  arbitrary  initial
superposition of $g$ states.  We  conclude that in general in the case
of $N_g, N_f<N_e$  it is impossible to remove  all the population from
the $g$ set in  a STIRAP-like population transfer process.  Exceptions
occur when the matrix  $\widetilde{\bm\Pi}$ is identically zero.  Then
the uncoupled MS states of the $e$ set are decoupled not only from the
$f$ set, but also from the $g$ set.

When  $N_e-N_f<N_g$ the  second  MS transformation  produces from  the
Hamiltonian of Eq.~(\ref{Hamb}) the matrix
\begin{equation}\label{Hambt}
   {\widetilde{{\bm H}}}'(t)=\left[\begin{array}{ccccc}
       {\bm 0}&{\bm 0}&p(t)\widetilde{{\bm P}}&{\bm 0}&{\bm 0}\\
       {\bm 0}&{\bm 0}&p(t)\widetilde{{\bm P}}'&p(t)\widetilde{{\bm\Pi}} &
       {\bm 0}\\
       p(t)\widetilde{{\bm P}}^\dagger&p(t)\widetilde{{\bm P}}^{\prime\dagger}
       &\hbar{\bm \Delta}&{\bm 0}&s(t)\widetilde{\bm \Sigma}\\
       {\bm 0}&p(t)\widetilde{{\bm\Pi}}^{\dag}&{\bm 0}&\hbar{\bm \Delta}&{\bm
         0}\\
       {\bm 0}&{\bm 0}&s(t)\widetilde{\bm \Sigma}^\dagger&{\bm
         0}&{\bm 0}\\
     \end{array}\right]\,.
\end{equation}
The situation $N_e-N_f\geq N_g$ can be treated similarly.  In order to
find  the  dark states  of  the  Hamiltonian  of Eq.~(\ref{Hambt})  we
proceed  in   the  same   way  as  in   Sec.~\ref{sec:piramid1}.   The
eigenvectors are parameterized as
\begin{equation}
   \label{vkparam2c}
   \widetilde{\advecb}_k \t=
   \left[\begin{array}{c}
       \widetilde{\bm x}_k \t \\
       \widetilde{\bm x}'_k \t \\
       \widetilde{\bm y}_k \t \\
       \widetilde{\bm y}'_k \t \\
       \widetilde{\bm z}_k \t\\
     \end{array}\right]\,.\quad
    \begin{array}{c}
      {g} \\ {g'} \\ {e} \\ {e'} \\ {f}
    \end{array}
\end{equation}
The  eigenvalue  equation is  defined  by Eq.~(\ref{eigeneqdef}).   By
inserting the Hamiltonian of (\ref{Hambt}) and the parameterization of
the  eigenvector Eq.~(\ref{vkparam2c}) into  Eq.~(\ref{eigeneqdef}) we
obtain  five groups  of coupled  linear equations  for $\widetilde{\bm
  x}_k$,     $\widetilde{\bm     x}'_k$,    $\widetilde{\bm     y}_k$,
$\widetilde{\bm y}'_k$, and $\widetilde{\bm z}_k$.  The $N_D$ linearly
independent  dark  eigenvectors $\widetilde{\advecb}_0^{(l)}(t)$  that
belong to the eigenvalue $\eigenv_0=0$, are given by
\begin{equation}
   \widetilde{\advecb}_0^{(l)}(t) = {1\over {\cal N}_0 \t}\left[
     \begin{array}{c}
       s(t)\widetilde{\bm x}_0   \\ s(t)\widetilde{\bm x}'_0  \\ {\bm
         0} \\ {\bm 0} \\
       -p(t)\widetilde{\bm \Sigma}^{-1}[\widetilde{{\bm
           P}}^\dagger\widetilde{\bm x}_0 +
       \widetilde{{\bm P}}^{\prime\dagger}\widetilde{\bm x}'_0]
     \end{array}\right]\,,
   \label{darkstates3}
\end{equation}
where, the  constant vector $\widetilde{\bm x}'_0$  should satisfy the
extra condition
\begin{equation}\label{zero-subspace}
   \widetilde{\bm \Pi}\widetilde{\bm x}'_0={\bm   0}\,.
\end{equation}
This condition says that in a dark state no population can be in those
$g$-set  states  that are  linked  to  uncoupled  $e$-set states.   An
example to this configuration is shown later in Sec.~\ref{subsec:121}.
The dimension $N_D$  of the dark subspace is  equal to $N_g-(N_e-N_f)$
plus the dimension of  the zero-subspace of the matrix $\widetilde{\bm
  \Pi}$,  Eq.~(\ref{zero-subspace}), where  we  assumed that  $N_g\geq
(N_e-N_f)$.  For  $N_g< (N_e-N_f)$ the dimension of  the dark subspace
is  equal  to  the  dimension  of  the  zero-subspace  of  the  matrix
$\widetilde{\bm \Pi}$.

It    is    useful   to    orthogonalize    the    dark   states    of
Eq.~(\ref{darkstates3}).   The  orthogonality  relation  is  given  by
Eq.~(\ref{ortho}).   In   this  case   we   find   that  the   vectors
$[\widetilde{\bm  x}^{(l)}_0 \, \widetilde{\bm  x}^{\prime (l)}_0]^T$,
$l=1\ldots N_D$ should be the eigenvectors of the Hermitian matrix
\begin{equation}\label{metric2}
   \left[\begin{array}{c}
       \widetilde{\bm P} \\ \widetilde{\bm P}'
     \end{array}\right]\widetilde{\bm \Sigma}^{-1\dagger}
   \widetilde{\bm  \Sigma}^{-1} [\widetilde{{\bm  P}}^\dagger
   \widetilde{{\bm  P}}^{\prime\dagger}]\,,
\end{equation}
with the restriction of Eq.~(\ref{zero-subspace}).

The other  eigenvectors, belonging to non-zero  eigenvalues, are given
by
\begin{equation}
   \widetilde{\advecb}_k \t = {1\over {\cal N}_k
     \t}\left[\begin{array}{c}
       p(t)\widetilde{{\bm P}}
       \widetilde{\bm y}_k \t\\ 
       p(t) \left[ \widetilde{{\bm P}}'
         \widetilde{\bm y}_k \t +
         \widetilde{{\bm \Pi}} \widetilde{\bm
           y}'_k \t \right]\\ 
       \eigenv_k \t \widetilde{\bm y}_k \t \\
       \eigenv_k \t \widetilde{\bm y}'_k \t \\ 
       s(t)\widetilde{\bm \Sigma}^{\dagger} \widetilde{\bm y}_k \t
     \end{array}\right],
   \label{eq:vk2}
\end{equation}
where  ${\cal  N}_k \t$  is  a  normalization  factor and  the  vector
$[\widetilde{\bm y}_k  \t\, \widetilde{\bm y}'_k  \t]^T$ satisfies the
eigen-equation
\begin{widetext}
   \begin{equation}
     \left[ \begin{array}{cc}
         p(t)^2(\widetilde{{\bm P}}^\dagger
         \widetilde{{\bm P}} +
         \widetilde{{\bm P}}^{\prime\dagger}
         \widetilde{{\bm P}}') +
         s(t)^2 \widetilde{\bm
           \Sigma}\widetilde{\bm \Sigma}^{\dagger} &
         p(t)^2
         \widetilde{{\bm P}}^{\prime\dagger} \widetilde{{\bm\Pi}} \\
         p(t)^2 \widetilde{{\bm\Pi}}^{\dag}\widetilde{{\bm P}}^{\prime}  &
         p(t)^2 \widetilde{{\bm\Pi}}^{\dag}\widetilde{{\bm \Pi}}
       \end{array}\right] \left[ \begin{array}{c}
         \widetilde{\bm
           y}_k \t\\
         \widetilde{\bm y}'_k \t \end{array}\right ]
     =   \eigenv_k \t [\eigenv_k\t-\hbar\Delta] \left[ \begin{array}{c}
         \widetilde{\bm y}_k \t \\ \widetilde{\bm y}'_k \t
       \end{array}\right ]\,.
     \label{eq:eigen2}
   \end{equation}
\end{widetext}
The  states  of Eq.~(\ref{eq:vk2})  are  bright  states, because  they
include components from the $e$ set.

The eigenvectors  of the Hamiltonian  ${\bm H}(t)$ in the  bare atomic
basis can be obtained as
\begin{equation}
   \advecb_k \t = {\bm U}^\dagger{\bm U}^{\prime\dagger} \left[
     \begin{array}{c}
       \widetilde{\bm x}_k \t \\
       \widetilde{\bm x}'_k \t  \\
       \widetilde{\bm y}_k \t \\
       \widetilde{\bm y}'_k \t  \\
       \widetilde{\bm z}_k \t \\
     \end{array} \right] =
   \left[ \begin{array}{c}
       {\bm A}^{\prime\dagger}\left[\begin{array}{c}
           \widetilde{\bm x}_k \t \\
           \widetilde{\bm x}'_k \t  \\
         \end{array}\right] \\ \\
       {\bm B}^{\dagger}\left[\begin{array}{c}
           \widetilde{\bm y}_k \t \\
           {\bm B}^{\prime\dagger}\widetilde{\bm y}'_k \t \\
         \end{array}\right] \\ \\
       {\bm A}^\dagger \widetilde{\bm z}_k \t
     \end{array} \right]\,.
\end{equation}

\section{Adiabaticity conditions and time evolution}\label{sec:adiab-tevol}

In  Sec.~\ref{sec:mstrafo} we have  presented the  dark states  of our
degenerate  system. Once  we  have  the dark  states  we may  consider
adiabatic evolution of the system  in the dark subspace. There are two
questions  that  should  be  addressed in  connection  with  adiabatic
evolution:
\begin{itemize}
\item[(1)] What are the conditions needed to ensure adiabatic evolution?
\item[(2)] If there are several degenerate dark states of a system, in
     general there are nonadiabatic couplings among them. How can we find
     the time evolution of the system in this case?
\end{itemize}

To answer  the first question we  apply the basic  theory of adiabatic
evolution  \cite{Messiah},   which  assures  that   the  evolution  is
adiabatic if any nonadiabatic couplings among the adiabatic states are
negligible compared with their  energy separation. In our model system
we have a dark subspace that is spanned by states that have eigenvalue
zero.  The  other adiabatic states,  the bright states,  have non-zero
eigenvalues.  Because we  want the state vector to  remain in the dark
subspace,  we require  that the  dark subspace  be separated  from the
bright one, as expressed by the condition
\begin{equation}\label{adi-cond}
      \hbar |\langle \widetilde\advec^{(l)}_0(t)
      |\dot{\widetilde\advec}_k(t)\rangle|\ll |\eigenv_k \t|\,,
\end{equation}
where $l=1\ldots N_D$ and $k=1\ldots N_B$, with $N_B$ being the number
of bright states. The dot  denotes time derivative. We may insert into
Eq.~(\ref{adi-cond})  any   set  of   dark  and  bright   states  from
Sec.~\ref{sec:mstrafo}.  For  example using  the dark states  given by
Eq.~(\ref{darkstates1}) and  the bright states  from Eq.~(\ref{eq:vk})
we find
\begin{equation}\label{adi-cond2}
    \frac{\hbar}{{\cal N}^{(l)}_0 \t{\cal N}_k \t}
    \left|s(t)\dot{p}(t)  - p(t)\dot{s}(t)
    \right|  \cdot \left |\langle {\bm x}^{(l)}_0|{\bm P}
      |{\bm y}_k \t\rangle\right| \ll  {|\eigenv_k \t|} \,.
\end{equation}
This  formula  closely resembles  the  adiabaticity  condition of  the
conventional nondegenerate three-level STIRAP \cite{AAMOP},
\begin{equation}
     \frac{1}{\Omega_0^2(t)}\left|S(t)\dot{P}(t)-P(t)\dot{S}(t)\right|\left|
       \begin{array}{c}
         \sin\varphi \\ \cos\varphi
       \end{array}\right|\ll \Omega_0(t)
     \left|\begin{array}{c}
       \cot\varphi \\ \tan\varphi
     \end{array}\right|\,,
\end{equation}
with $\Omega_0(t)$ and $\varphi$ defined as
\begin{equation}
    \Omega_0(t)=\sqrt{P^2(t)+S^2(t)}\,,\qquad \tan2\varphi=\frac{\Omega_0(t)}
    {\Delta}\,,
\end{equation}
respectively.  The  second factor on the  lhs of Eq.~(\ref{adi-cond2})
contains the  time-derivatives of the  envelope functions of  the pump
and Stokes  pulses, whereas the  third factor involves the  element of
the matrix ${\bm P}$ between the  $l$th dark state at the initial time
and  the excited-state-amplitudes of  the $k$th  bright state  at time
$t$.  On  the rhs $\eigenv_k$  is the eigenenergy associated  with the
$k$th   bright   state.     Whenever   the   adiabaticity   conditions
Eq.~(\ref{adi-cond2}) are fulfilled for  all dark and bright states of
the system, then the dark and bright subspaces evolve independently.

There remains the  task of determining the time  evolution in the dark
subspace.   When there are  several degenerate  dark states  there are
usually  nonadiabatic couplings  among them.   In case  of  the tripod
system \cite{Unanyan98,Theuer99},  due to  the special choice  for the
time-dependence  of  the  Stokes  pulses,  the  two  dark  states  mix
throughout the  population transfer process.  If the  dimension of the
dark subspace  is larger than two,  then in general there  is no exact
analytic solution  \cite{Kis01,Kis02}.  In  our case the  situation is
much simpler.  The nonadiabatic coupling between a pair of dark states
is        $\langle         \widetilde\advec^{(l)}_0        \t        |
\dot{\widetilde\advec}^{(k)}_0  \t  \rangle\,$.   By  evaluating  this
expression     for     any     pair     of    dark     states     from
Sec.~\ref{sec:stokes-mstrafo} we  always get identically  zero.  Hence
the  dark states do  not mix  throughout the  whole transfer  process.
This property simplifies the calculations considerably, since the time
evolution operator in the dark subspace is given by
\begin{equation}\label{op-tevol}
   {\bm U}(t,t_0)=\sum_{l=1}^{N_D} |\advec^{(l)}_0(t)\rangle \langle
   \advec^{(l)}_0 (t_0)|\,.
\end{equation}
For example, for $N_g \leq  N_e \leq N_f$ the vectors $\advecb^{(l)}_0
(t_0)$  are equal to  ${\bm x}_0^{(l)}$  from Eqs.~(\ref{darkstates1})
and  (\ref{ortho}).    Once  we   define  the  density   matrix  ${\bm
   \varrho}(t_0)$ of the system at time $t_0$ the density matrix at any
later time is given by
\begin{equation}\label{rho-tevol}
   {\bm\varrho}(t)={\bm U}(t,t_0){\bm\varrho}(t_0){\bm U}^{\dag}(t,t_0)\,,
\end{equation}
provided  that the  adiabaticity conditions  Eq.~(\ref{adi-cond2}) are
satisfied.    Note  that  this   formula  is   valid  {\em   only}  if
${\bm\varrho}(t_0)$ lies  entirely in  the dark subspace.   It follows
from Eqs.~(\ref{op-tevol})  and (\ref{rho-tevol}) that  {\em any} pure
or mixed  initial state of the  system occupying the  dark subspace of
the  $g$ set  is  transferred to  the $f$  set  in the  course of  the
population  transfer process.   In the  case of  a pure  initial state
${\Psi}(t_0)$ we have
\begin{equation}
   {\Psi}(t)={\bm U}(t,t_0){\Psi}(t_0)\,.
\end{equation}
Here  again, ${\Psi}(t_0)$  must have  components solely  in  the dark
subspace.  (Were  the state vector to have components  initially in the
bright  subspace,   then  adiabatic  evolution   would  maintain  such
presence.   Because the excited  states undergo  spontaneous emission,
their populations have the potential to interrupt the coherence of the
dynamics and thereby to diminish the population transfer.)

\section{Some examples}\label{sec:examples}

In this section we demonstrate  through some examples the usage of our
method.   To be  specific, we  consider atomic  transitions  where the
origin of the  degeneracy is the set of  degenerate magnetic sublevels
of angular  momentum states  in the absence  of a magnetic  field. Our
purpose is  to present some  typical configurations that may  occur in
realistic situations.

\subsection{The $J=1\leftrightarrow2\leftrightarrow3$ linkage}
\label{subsec:123}

\begin{figure}
  \includegraphics[width=7cm]{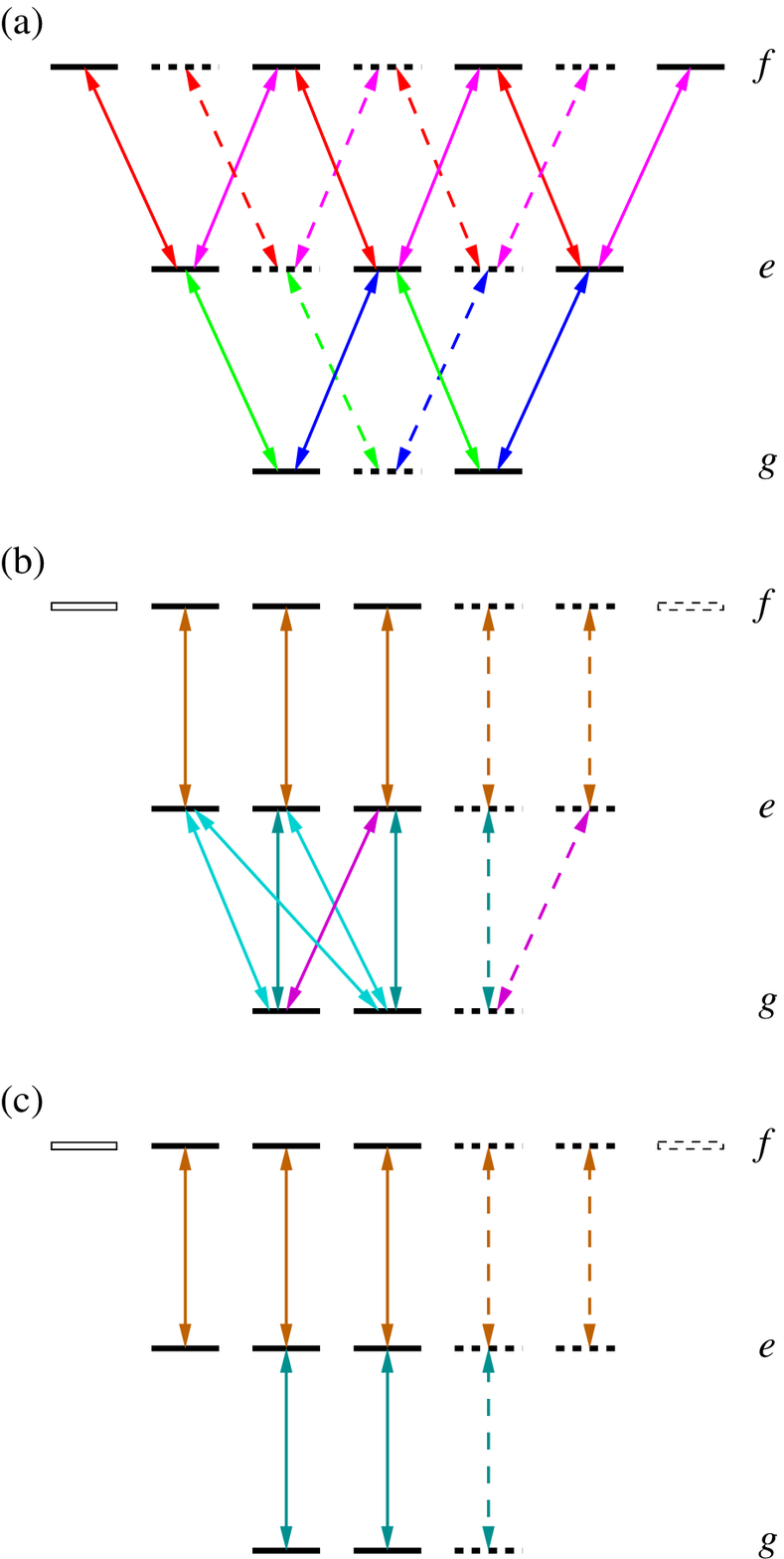} 
  \caption{(Color Online)
    The        coupling        configuration}       for        the
  $J=1\leftrightarrow2\leftrightarrow3$    linkage    with    only
  $\sigma^{\pm}$ polarized coupling  fields.  The system separates
  into two  independent subsystems; the smaller one  is shown with
  dashed lines, the larger one  with solid lines.  Frame (b) shows
  the  result of  the Stokes-field  MS transformation.   Frame (c)
  shows the  redefinition of the states  in the $g$,  $e$, and $f$
  sets according to Eq.~(\ref{sets}).
\label{fig:123scheme}
\end{figure}

In  this  example  we   consider  the  linkage  $J=1  \leftrightarrow2
\leftrightarrow3$, shown  in Fig.~\ref{fig:123scheme} and  assume that
only $\sigma^{\pm}$  fields are present.   In this case there  are two
independent  coupled systems:  the one  with $M_g=0$,  $M_e=\pm1$, and
$M_f=0,\pm2$  (shown  as  dashed   lines);  and  the  other  one  with
$M_g=\pm1$, $M_e=0,\pm2$, and $M_f=\pm1,  \pm2$ (shown as full lines).
The first one  has been studied in ref.~\cite{Kis03},  hence we do not
consider it  here.  For the  second, larger system, the  pump coupling
matrix $\bm P$ is given by
\begin{equation}\label{P234}
    \bm P = \frac{\hbar}{2}\sqrtfrac{1}{3}\left[ \begin{array}{ccc}
        \Omega_P^{(-)}&\sqrtfrac{1}{6}\Omega_P^{(+)}&0
        \vspace{5pt}  \\
        0&\sqrtfrac{1}{6}\Omega_P^{(-)}&\Omega_P^{(+)}
      \end{array}\right]\,,
\end{equation}
whereas the Stokes coupling matrix reads
\begin{equation}\label{S234}
   {\bm S} =
   \frac{\hbar}{2}\sqrtfrac{1}{5}\left[
     \begin{array}{cccc}
       \vspace{5pt}
       \Omega_S^{(-)}&\sqrtfrac{1}{15}\Omega_S^{(+)}&0&0
       \\
       \vspace{5pt}
       0&\sqrtfrac{2}{5}\Omega_S^{(-)}&\sqrtfrac{2}{5}\Omega_S^{(+)}&0
       \\
       0&0&\sqrtfrac{1}{15}
       \Omega_S^{(-)}&\Omega_S^{(+)}
     \end{array}\right].
\end{equation}
The  numeric   factors  in  front  of  the   $\Omega$-s  describe  the
Clebsch-Gordan coefficients. The Rabi frequencies are parameterized as
\begin{subequations}\label{123rabi}
   \begin{eqnarray}
     \Omega_P^{(+)}&=&\Omega_P \,e^{i\phi_P} \cos\eta\,, \\
     \Omega_P^{(-)}&=&\Omega_P \,e^{i\psi_P} \sin\eta\,, \\
     \Omega_S^{(+)}&=&\Omega_S \,e^{i\phi_S} \cos\theta\,, \\
     \Omega_S^{(-)}&=&\Omega_S \,e^{i\psi_S} \sin\theta\,,
   \end{eqnarray}
\end{subequations}
where the amplitudes $\Omega_{P,S}$ are nonnegative. The angles $\eta$
and  $\theta$ characterize  the pump  and Stokes  field polarizations,
respectively.

Here  $N_g   <  N_e  <   N_f$  ($2<3<4$),  hence  the   derivation  in
Sec~\ref{sec:piramid1} can  be applied.  As  a first step, we  have to
perform   the  Stokes  field   MS  transformation.    The  eigenvalues
$\lambda_k$ of  the matrix  ${\bm S}{\bm S}^{\dag}$  are given  by the
roots of a cubic equation, see Eq.~(\ref{eigvals123MS}).
\begin{figure}
   \includegraphics[width=8cm]{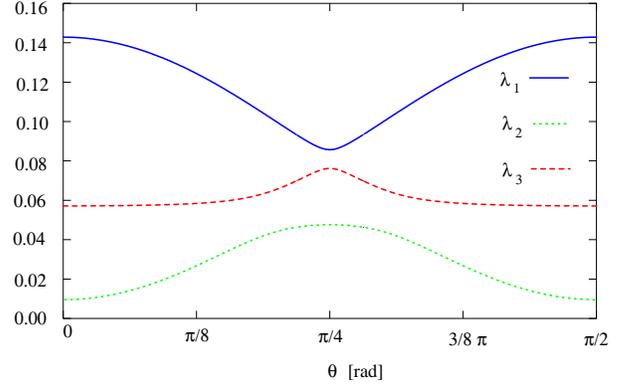} 
   \caption{ (Color Online) The eigenvalues of the matrix $\bm S\bm S^{\dag}$,
     Eq.~(\ref{S234}),  as  a function  of  the  polarization of  the
     Stokes  field.  The  eigenvalues  are measured  in  the units  of
     $(\hbar\Omega_S)^2$. }
   \label{fig:lambda_k}
\end{figure}
We  display  them in  Fig.~\ref{fig:lambda_k}  as  a  function of  the
polarization angle $\theta$.  They  are never zero, hence the complete
adiabatic population  transfer is possible for  {\em any} polarization
of  the  Stokes  field.   However,  their  amplitudes  depend  on  the
polarization,    which    affects    the    adiabaticity    conditions
Eqs.~(\ref{adi-cond})  and  (\ref{adi-cond2}).   The Stokes  field  MS
transformation matrices $\bm A$  and $\bm B$, Eq.~(\ref{Udef}), can be
calculated  in  a  straightforward  manner;  they  are  shown  in  the
Appendix~\ref{sec:123MS}.   Since  $N_f=N_e+1$,  the Stokes  field  MS
transformation  yields a  transformed coupling  matrix $\widetilde{\bm
  S}$ in the  form of the first row  in Eq.~(\ref{Str}).  The diagonal
part $\widetilde{\bm \Sigma}$ is given by
\begin{equation}\label{sigma234}
    \widetilde{\bm\Sigma} =
    \sqrtfrac{7}{20}\hbar\Omega_S\left[\begin{array}{ccc}
        \sqrt{\lambda_1}&0&0\\
        0&\sqrt{\lambda_2}&0\\
        0&0&\sqrt{\lambda_3}
      \end{array}\right]\,.
\end{equation}
There are $2+4-3=3$ dark states in this system: one is in the $f$ set,
an uncoupled state.  The space of $g$ is two-dimensional, $N_g=2$, and
hence    there    are    two    dark-states    in    the    form    of
Eq.~(\ref{darkstates1}).   The  vectors  ${\bm x}_0^{(k)}$  associated
with  these two  dark states  are  the eigenvectors  of the  Hermitian
matrix  ${\bm   M}$  of  Eq.~(\ref{metric})  and  are   given  in  the
Appendix~\ref{sec:123MS}.   The  two   dark  states  are  obtained  by
inserting  their structure  into Eq.~(\ref{darkstates1})  or,  for the
bare atomic basis, into Eq.~(\ref{advecb}).

\begin{figure}

\includegraphics[width=8cm]{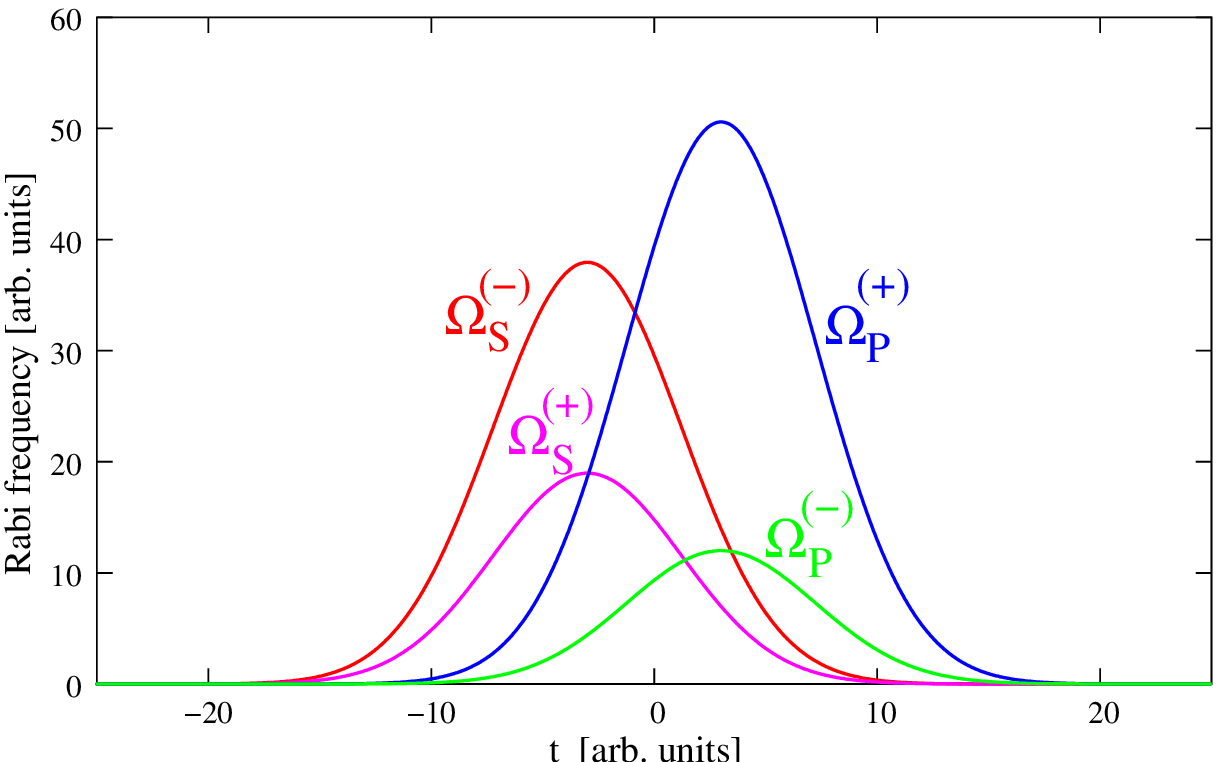}\\[10pt] 

\includegraphics[width=8cm]{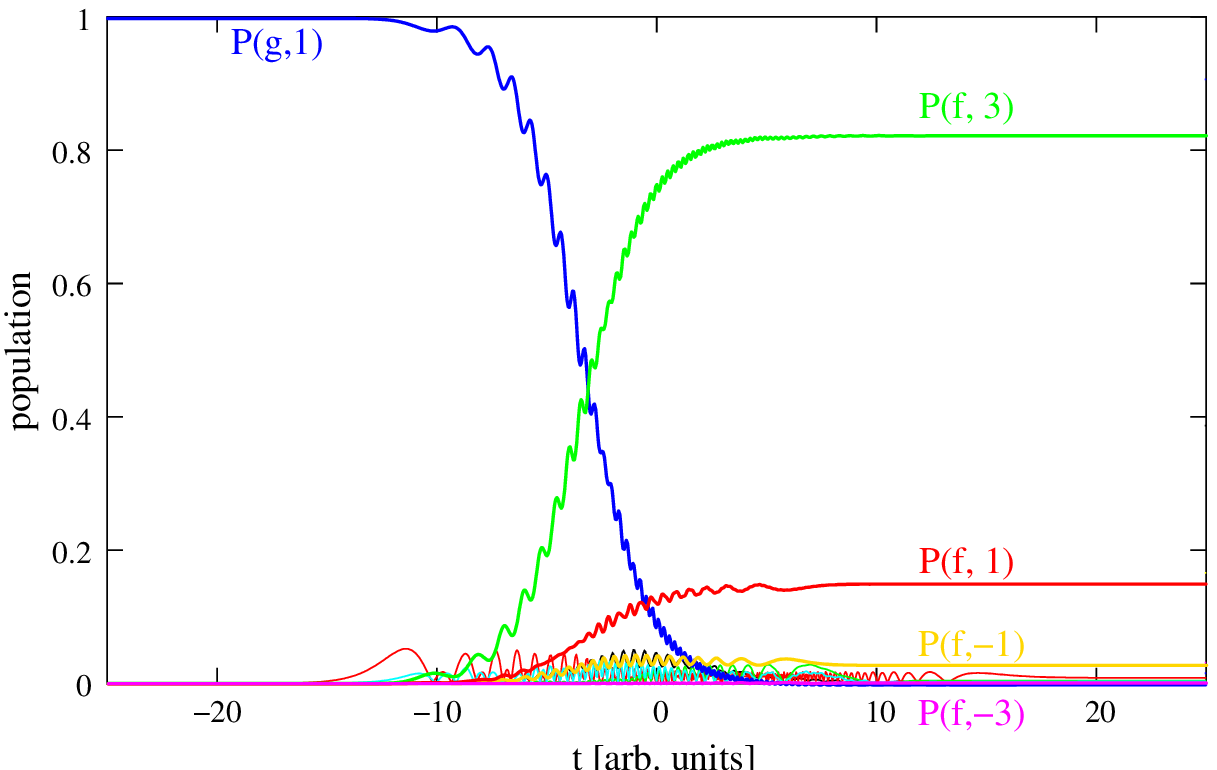} 
     \caption{ (Color Online) Upper frame:  the  pulse  sequence  used
       for  the  population transfer  process  in  the coupled  angular
       momentum   system  $J=1\leftrightarrow   2\leftrightarrow   3$.
       Initially only the  state $|g,J_g=1,M_g=1\rangle$ was populated.
       Lower frame: The population evolution. }
\label{fig:123pop}
\end{figure}

We have performed  numerical simulations to check the  validity of our
analytic results. In Fig.~\ref{fig:123pop} initially the system was in
the state $|g,J_g=1,M_g=1\rangle$.  The envelope functions of the pump
and  Stokes pulses, respectively,  are $p(t)=  \exp(-[t-3]^2/6^2)$ and
$s(t)=  \exp(-[t+3]^2/6^2)$;  and  $\Omega_P=52$, $\Omega_S=42$.   The
polarizations   are   characterized    by   $\eta=1.3376$   rad,   and
$\theta=0.4636$   rad.    The    phases   are   chosen   randomly   as
$\phi_P=1.1814$   rad,  $\psi_P=0$   rad,  $\phi_S=1.8925$   rad,  and
$\psi_S=2.8198$ rad.  The  detuning $\Delta$ is set to  zero.  We have
found again very good  agreement between the analytic calculations and
the numeric simulation.

\begin{figure}
   \includegraphics[width=8cm]{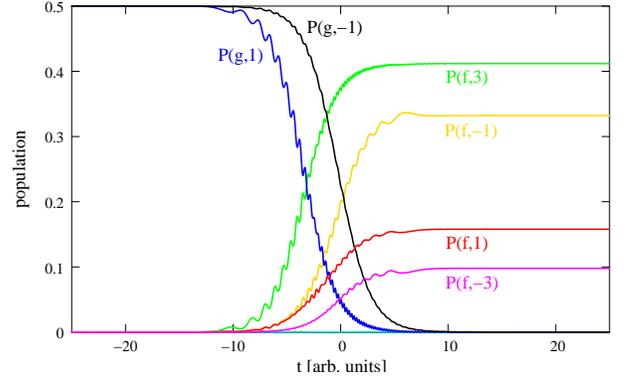} 
   \caption{ (Color Online) Same as Fig.~\ref{fig:123pop}, but for a
     mixed   initial  state,   $P_i(1,-1)=P_i(1,1)=1/2$,   all  initial
     coherences are zero.  }
   \label{fig:123-kevert-pop}
\end{figure}

We also  considered a mixed initial  state, when the  initial state of
the system  is chosen as half of  the population is placed  on each of
the  $|g,J_g=1,M_g=\pm1\rangle$  states,  and  the coherence  is  zero
between  them.  The numerically  calculated dynamics  is shown  in
Fig.~\ref{fig:123-kevert-pop}. We  can see  that despite of  the mixed
initial state, the complete population can be transferred from the $g$
set to the $f$ set. The pulse  sequence is the same as in the previous
example.

\subsection{The $J=1\leftrightarrow1\leftrightarrow1$ linkage}
\label{subsec:111}

\begin{figure}
      \includegraphics[width=3.5cm]{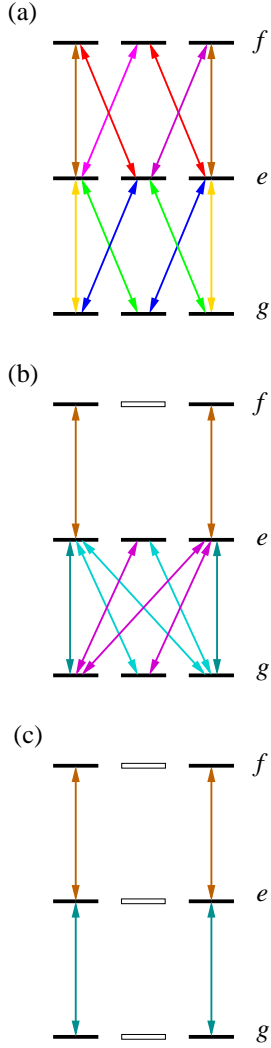} 
      \caption{(Color Online) Same as Fig.~\ref{fig:123scheme} for
        equal    state-degeneracies.      For    equal    $J$-s,    the
        $M=0\leftrightarrow 0$  transition is dipole-forbidden,  and we
        cannot select  a basis such that there  occur couplings between
        all pairs  of states of the degenerate  sets.  This restriction
        results from  the property of the  Clebsch-Gordan coefficients.
        Frame (c) shows the redefinition of the states in the $g$, $e$,
        and  $f$ sets  according  to Eq.~(\ref{sets}),  leading to  two
        independent three-state linkages. }
\label{fig:111scheme}
\end{figure}

As      another     example      we      consider     the      linkage
$J=1\leftrightarrow1\leftrightarrow1$             shown             in
Fig.~\ref{fig:111scheme}.  In  this case $N_g=N_e=N_f$,  and hence the
derivation  in  Sec.~\ref{sec:piramid1}  is  applicable.   This  is  a
counter-example  to the  general condition  of  Eq.~(\ref{cond}): even
though the  condition Eq.~(\ref{cond}) is satisfied, in  this case the
complete removal of an  arbitrary population distribution from the $g$
set is impossible in the STIRAP way.

The  coupling  matrices  $\bm  S$  and  $\bm  P$  in  the  Hamiltonian
Eq.~(\ref{ham}) are given by
\begin{equation}
    \bm X = \frac{\hbar}{2}
    \frac{1}{\sqrt{6}}
    \left[ \begin{array}{ccc}
        \vspace{5pt}
        -\Omega_X^{(\pi)}  &-\Omega_X^{(+)}  &0
        \\  \vspace{5pt} [5pt]
        \Omega_X^{(-)}&0&-\Omega_X^{(+)}
        \\ [5pt]
       0&
       \Omega_X^{(-)}&\Omega_X^{(\pi)}
     \end{array}\right],
\end{equation}
for $\bm X=\bm  S$ or $\bm P$.  The factor  $1/\sqrt{6}$ and the $\pm$
signs describe the  Clebsch-Gordan coefficients.  The Rabi frequencies
$\Omega_X^{(\pm,\pi)}$  correspond to the  $\sigma^{+}$, $\sigma^{-}$,
and  $\pi$ polarizations,  respectively.  Note  that a  selection rule
nullifies transitions $M = 0 \leftrightarrow M = 0$.

As   described  in   Sec.~\ref{sec:stokes-mstrafo},  we   perform  the
Stokes-field  MS  transformation to  diagonalize  the Stokes  coupling
matrix  ${\bm  S}$.   The  eigenvalues  of  the  matrix  ${\bm  S}{\bm
   S}^{\dag}$ provide  the squared moduli  of the diagonal  elements of
the  matrix   $\widetilde{\bm  \Sigma}=\widetilde{\bm  S}={\bm  B}{\bm
   S}{\bm A}^{\dag}$, Eq.~(\ref{trafo}). They are given by
\begin{equation}
    0,\quad   \pm\frac{1}{6}\left(\Omega_{S}^{({\rm rms})}\right)^2\,,
\end{equation}
with  $\left(\Omega_{S}^{({\rm  rms})}\right)^2=|\Omega_S^{(+)}|^2  +
|\Omega_S^{(-)}|^2 + |\Omega_S^{(\pi)}|^2$.  One of the eigenvalues is
always zero and therefore, although the system satisfies the condition
for  complete  population transfer,  Eq.~(\ref{cond}),  the null  Rabi
frequency prevents complete transfer.

\begin{figure}

\includegraphics[width=8cm]{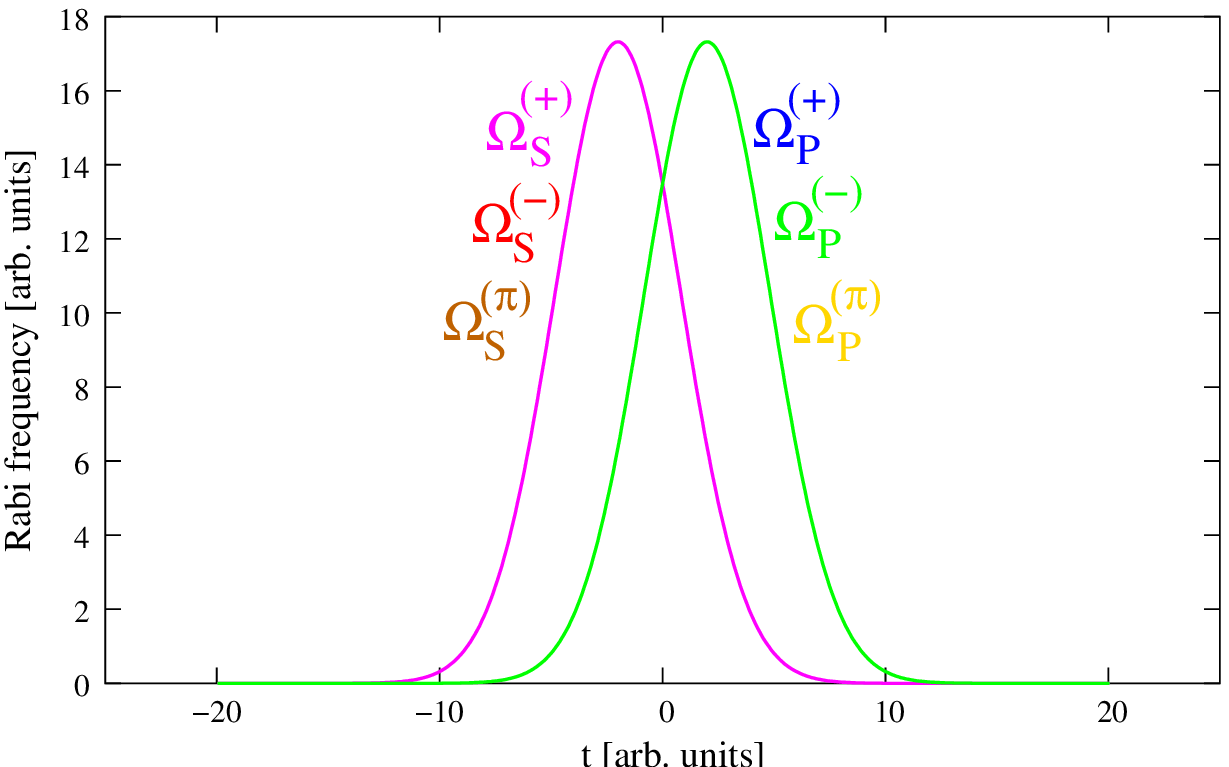}\\[10pt] 

\includegraphics[width=8cm]{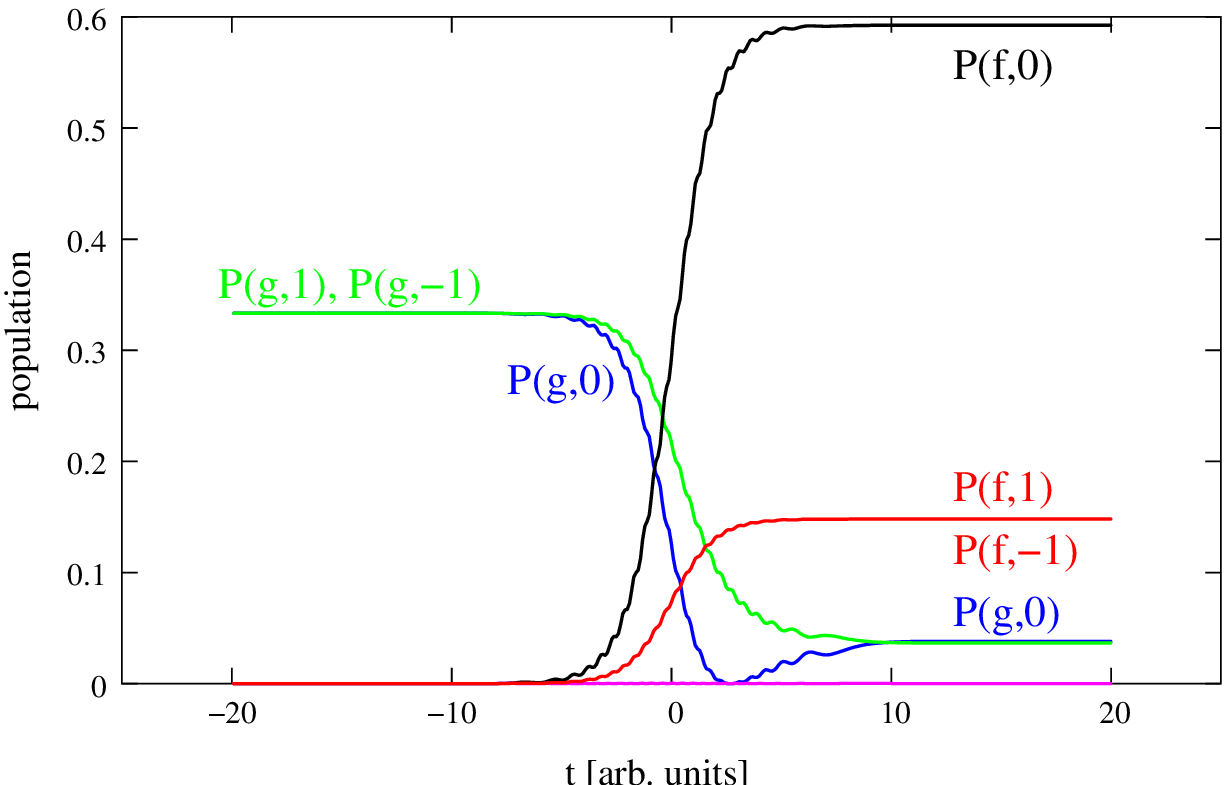} 
     \caption{ (Color Online) Upper frame:  the  pulse  sequence  used
       for  the  population transfer  process  in  the coupled  angular
       momentum   system  $J=1\leftrightarrow   1\leftrightarrow   1$.
       Initially        the       state       $(|g,J_g=1,M_g=-1\rangle-
       |g,J_g=1,M_g=0\rangle    +|g,J_g=1,M_g=1\rangle)/\sqrt{3}$   was
       populated.  Lower frame: The  population evolution.  Part of the
       population is left  in the $g$ set, because some  of the MS Rabi
       frequencies vanish. }
\label{fig:111-evol}
\end{figure}

Fig.~\ref{fig:111-evol} demonstrates  the population transfer  in this
system.      Initially    the    system     was    in     the    state
$(|g,J_g=1,M_g=-1\rangle-            |g,J_g=1,M_g=0\rangle           +
|g,J_g=1,M_g=1\rangle)/ \sqrt{3}$.  The envelope functions of the pump
and     Stokes     pulses,     respectively,     are     chosen     as
$p(t)=\exp(-[t-2]^2/4^2)$     and    $s(t)=\exp(-[t+2]^2/4^2)$,    and
$\Omega_P=\Omega_S=30$.  The intensity  is equally distributed among the
$\sigma^{+}$,  $\sigma^{-}$,  and  $\pi$  components of  the  exciting
fields.   The  detuning  $\Delta$  is  set to  zero.   We  have  found
excellent agreement between the  analytic calculations and the numeric
simulation.   The adiabaticity conditions,  Eq.~(\ref{adi-cond2}), are
also fulfilled throughout the relevant part of the population transfer
process.

\subsection{The $J=1\leftrightarrow 2 \leftrightarrow 1$ linkage}
\label{subsec:121}

\begin{figure}
     \includegraphics[width=5cm]{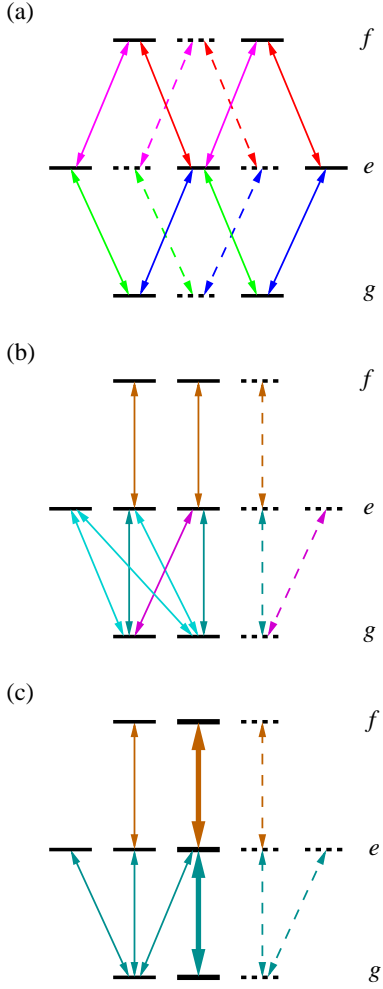} 
     \caption{(Color Online) Same as Fig.~\ref{fig:123scheme} for the
       $J=1\leftrightarrow2\leftrightarrow1$    linkage    with    only
       $\sigma^{\pm}$ polarized coupling  fields.  The system separates
       into two  independent subsystems; The smaller one  is shown with
       dashed lines, the larger one  with solid lines.  Frame (b) shows
       the  result of  the Stokes-field  MS transformation.   Frame (c)
       shows the result  of the pump field MS  transformation {\em and}
       redefinition  of  the states  in  the  $g$,  $e$, and  $f$  sets
       according to  Eq.~(\ref{sets}).  In  general, there is  one dark
       state  in  the  larger  system  which  is  associated  with  the
       three-state linkage indicated by heavy lines. }
\label{fig:121scheme}
\end{figure}

In  our last  example we  consider the  linkage  $J=1 \leftrightarrow2
\leftrightarrow1$, shown  in Fig.~\ref{fig:121scheme} and  assume that
only $\sigma^{\pm}$  fields are present.   In this case there  are two
independent  coupled systems:  the one  with $M_g=0$,  $M_e=\pm1$, and
$M_f=0$    (shown   as   dashed    lines)   discussed    recently   in
ref.~\cite{Shah02}, and  the other one  with $M_g=\pm1$, $M_e=0,\pm2$,
and  $M_f=\pm1$  (shown  as  full  lines).  This  is  a  twin  diamond
configuration.  For  the larger system, the pump  coupling matrix $\bm
P$ is given by
\begin{equation}
   {\bm P} = \frac{\hbar}{2}\left[ \begin{array}{ccc}
       \sqrtfrac{1}{3}\Omega_P^{(-)}&\sqrtfrac{1}{18}\Omega_P^{(+)}&0\\
       \vspace{5pt}
       0&\sqrtfrac{1}{18}\Omega_P^{(-)}&\sqrtfrac{1}{3}\Omega_P^{(+)}
     \end{array}\right]\,.
\end{equation}
whereas the Stokes coupling matrix reads
\begin{equation}
   {\bm S} = \frac{\hbar}{2}\left[ \begin{array}{cc}
       \frac{\sqrt{3}}{5}\Omega_S^{(-)}&0\\
       \vspace{5pt}
       \sqrtfrac{1}{50}\Omega_S^{(+)}&\sqrtfrac{1}{50}\Omega_S^{(+)}\\
       \vspace{5pt}
       0&\frac{\sqrt{3}}{5}\Omega_S^{(-)}
     \end{array}\right]\,.
\end{equation}
The  parameterization of  the Rabi  frequencies  $\Omega_X^{(\pm)}$ is
given by Eq.~(\ref{123rabi}).  Here  $N_g, N_f < N_e$ ($2,2<3$), hence
the derivation in Sec~\ref{sec:diamond} is applicable. The sequence of
the dimension of the  subspaces violate the condition $N_g\leq N_e\leq
N_f$, therefore, in general a STIRAP-like complete population transfer
is  not possible.   However, this  is another  counter-example  to the
general  condition  of  Eq.~(\ref{cond}):  even though  the  condition
Eq.~(\ref{cond}) is violated, we show  that the complete removal of an
arbitrary population distribution from the  $g$ set is possible in the
STIRAP way for a special choice of pulse polarizations and phases.
 
As  usual, we  start with  the  Stokes field  MS transformation.   The
eigenvalues of  the matrix  ${\bm S}{\bm S}^{\dag}$  are given  by the
roots of a quadratic equation, which read
\begin{equation}\label{121seigs}
   \lambda_{1,2}=\frac{7}{100}\pm\frac{1}{100}\sqrt{24\cos^22\theta+1}\,.
\end{equation}
The  Stokes field  MS transformation  matrices  $\bm A$  and $\bm  B$,
Eq.~(\ref{Udef}), can be calculated in a straight forward manner, they
are  shown in  the Appendix~\ref{sec:121MS}.   Since  $N_f=N_e-1$, the
Stokes field  MS transformation  yields a transformed  coupling matrix
$\widetilde{\bm S}$  in the form of  the last row  in Eq.~(\ref{Str}). 
The diagonal part $\widetilde{\bm \Sigma}$ is given by
\begin{equation}
   \widetilde{\bm\Sigma} = \frac{\hbar}{2} \Omega_S\left[\begin{array}{cc}
       \sqrt{\lambda_1}&0\\
       0&\sqrt{\lambda_2}
     \end{array}\right]\,.
\end{equation}
The  eigenvalues of  Eq.~(\ref{121seigs}) are  always  positive, hence
this matrix  is nonsingular for  {\em any} polarization of  the Stokes
field.  The  Stokes field MS transformation yields  two $e-f$ linkages
and  an  $e$  state  which  is  not coupled  to  any  $f$  state,  see
Fig.~\ref{fig:121scheme}b.    Now,   following   the   derivation   of
Sec~\ref{sec:diamond} we  perform a  second MS transformation  for the
pump field.  The transformation  matrix is given by Eq.~(\ref{Updef}). 
In our case, the $2\times 2$ unitary matrix ${\bm A}'$ is defined as
\begin{widetext}
\begin{equation}
   {\bm A}'
   =\left[\begin{array}{cc}
       \cos\theta&e^{-i(\psi_S-\phi_S)}\sin\theta\\
       e^{-\frac12 i(\phi_S-\psi_S+\phi_P+\psi_P)}\sin\theta&
       -e^{-\frac12 i(\psi_S-\phi_S+\phi_P+\psi_P)}\cos\theta
     \end{array}
   \right]\,,
\end{equation}
while the  matrix ${\bm  B}'$ is a  scalar now,  and chosen as  unity. 
Since in this case $N_e-N_f<N_g$ ($3-2<2$), the transformed pump field
coupling  matrix  takes  the  form  of  Eq.~(\ref{pi1}).   The  matrix
$\widetilde{\bm \Pi}$ is a scalar, that reads
\begin{equation}
   \widetilde{\bm \Pi}=\begin{array}{c}
       -\frac{\Omega_P}{2\sqrt{3}\sqrt{2-\cos^2 2\theta}}\left(
         \cos\eta\cos\theta e^{\frac12i(\psi_S-\phi_S+\phi_P-\psi_P)}
         -\sin\eta\sin\theta e^{-\frac12i(\psi_S-\phi_S+\phi_P-\psi_P)}\right)
     \end{array}\,.
\end{equation}
\end{widetext}
This is nonzero  in general, hence one of the $g$  states is linked to
the {\em uncoupled}  $e$ state. Therefore, there is  one dark state in
the system, which reads
\begin{equation}\label{dark1}
   \widetilde{\bm \Phi}_0^{(1)}(t)=\frac{1}{{\cal N}_0^{(1)}(t)}
   \left[\begin{array}{c}
       s(t)\\
       0\\
       0\\
       0\\
       0\\
       -p(t)\widetilde{\bm \Sigma}^{-1}\bm B_a \bm P^\dagger {\bm A'}^\dagger
       \left[
         \begin{array}{c}1\\0\end{array}
       \right]
     \end{array}
   \right]\,.
\end{equation}
This  dark state  is associated  with the  three-state linkage  in the
middle of Fig.~\ref{fig:121scheme}c, indicated by heavy lines.

However, for
\begin{subequations}\label{cptdi2}
\begin{eqnarray}
   \psi_S-\phi_S+\phi_P-\psi_P&=&k\pi\,,\\
   \theta+(-1)^k\eta=\frac12\pi\,,
\end{eqnarray}
\end{subequations}
where $k$ is an integer, the scalar $\widetilde{\bm \Pi}$ vanishes. As
a result, the {\em uncoupled} $e$ state becomes decoupled from the $g$
state as well.  Therefore,  beside the dark state of Eq.~(\ref{dark1})
there is an other one
\begin{equation}
   \widetilde{\bm \Phi}_0^{(2)}(t)=\frac{1}{{\cal N}_0^{(2)}(t)}
   \left[\begin{array}{c}
       0\\
       s(t)\\
       0\\
       0\\
       0\\
       -p(t)\widetilde{\bm \Sigma}^{-1}\bm B_a \bm P^\dagger {\bm A'}^\dagger
       \left[
         \begin{array}{c} 0\\1 \end{array}
       \right]
     \end{array}
   \right]\,.
\end{equation}
In summary:  {\em complete} population  transfer is possible  from the
$g$ set to  the $f$ set for the special  choice of pulse polarizations
and  phases Eq.~(\ref{cptdi2}).   It  is important  to  note that  the
condition  for complete transfer  Eq.~(\ref{cptdi2}) is  equivalent to
that  for   the  diamond  configuration   \cite{Shah02}.   Hence,  the
population from the total $g$ set  can be transferred into the $f$ set
if the condition Eq.~(\ref{cptdi2}) is fulfilled.

\begin{figure}

\includegraphics[width=8cm]{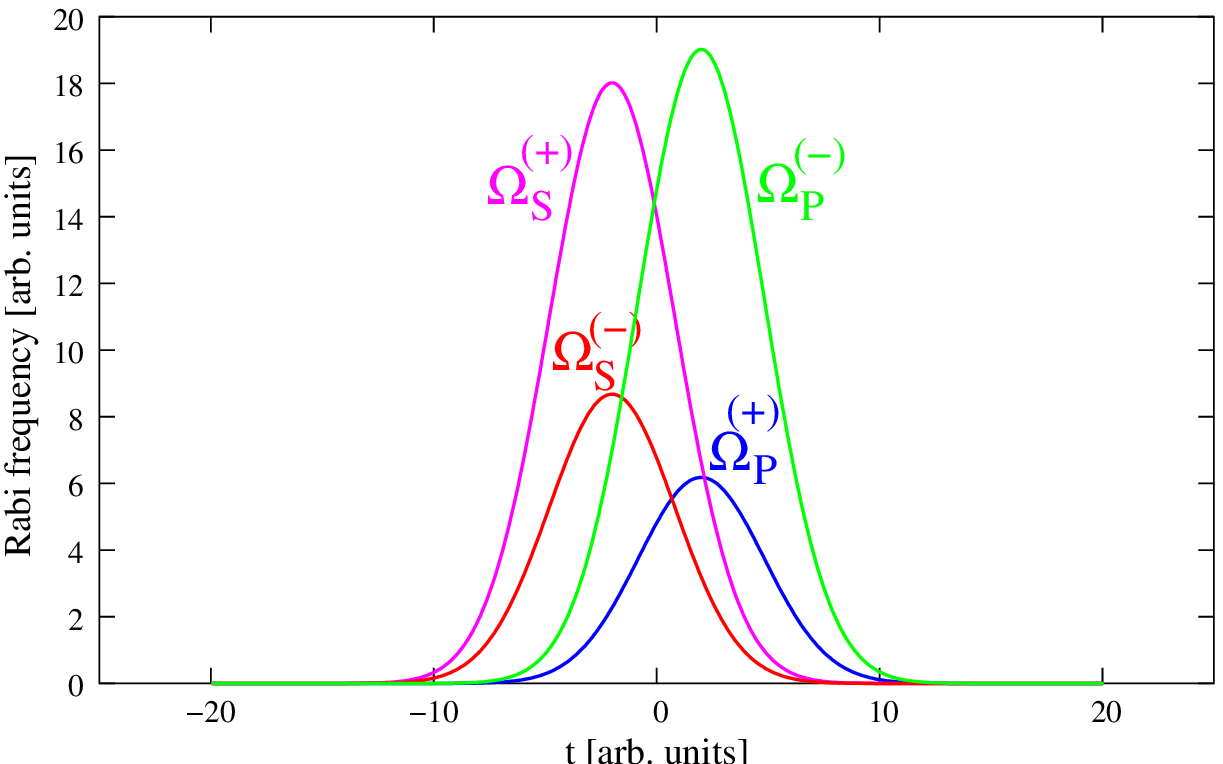}\\[10pt] 

\includegraphics[width=8cm]{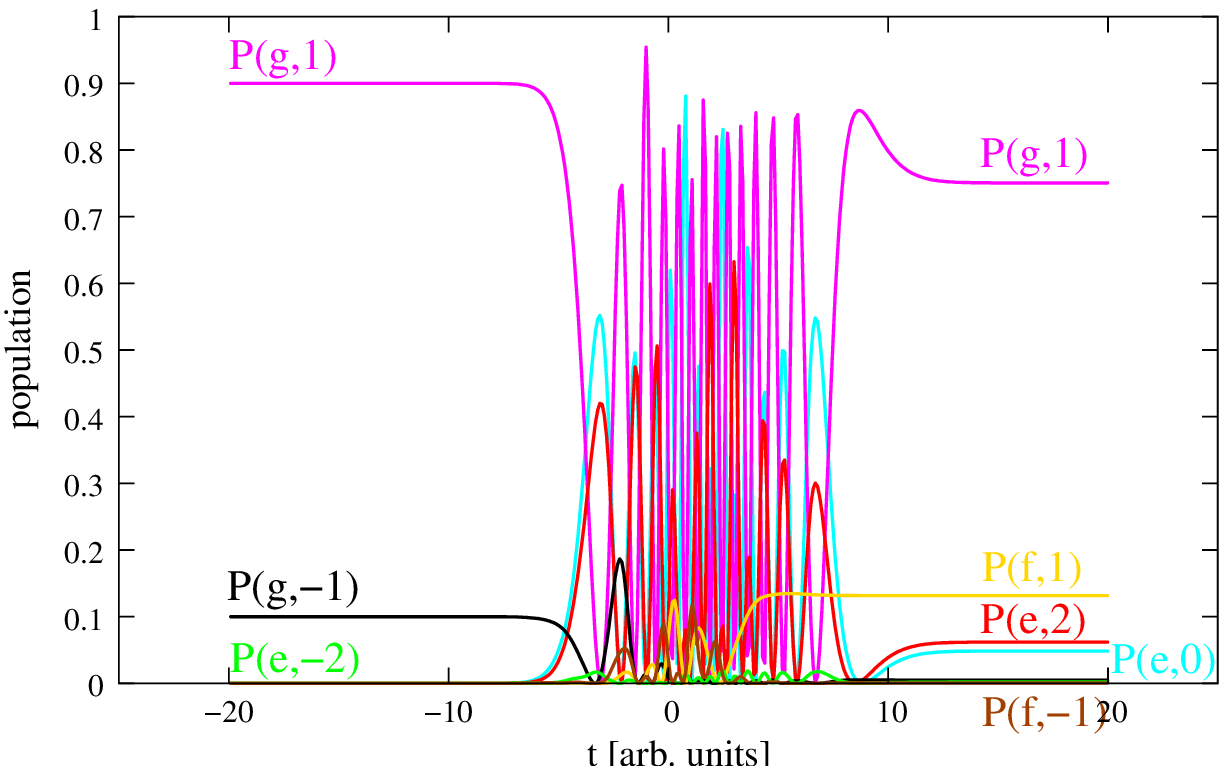} 
     \caption{ (Color Online) Upper frame:  the  pulse  sequence  used
       for  the  population transfer  process  in  the coupled  angular
       momentum   system  $J=1\leftrightarrow   2\leftrightarrow   1$.
       Initially   the   state  $\cos(\alpha)   |g,J_g=1,M_g=-1\rangle+
       \sin(\alpha)  |g,J_g=1,M_g=1\rangle$  with $\alpha=\arctan(1/3)$
       was  populated.   Lower frame:  The  population evolution.   The
       pulses  are chosen so  that the  special condition  for complete
       transfer  Eq.~(\ref{cptdi2})  is  violated,  hence part  of  the
       population is left in the $g$ and $e$ sets. }
\label{fig:121A-evol}
\end{figure}

\begin{figure}

\includegraphics[width=8cm]{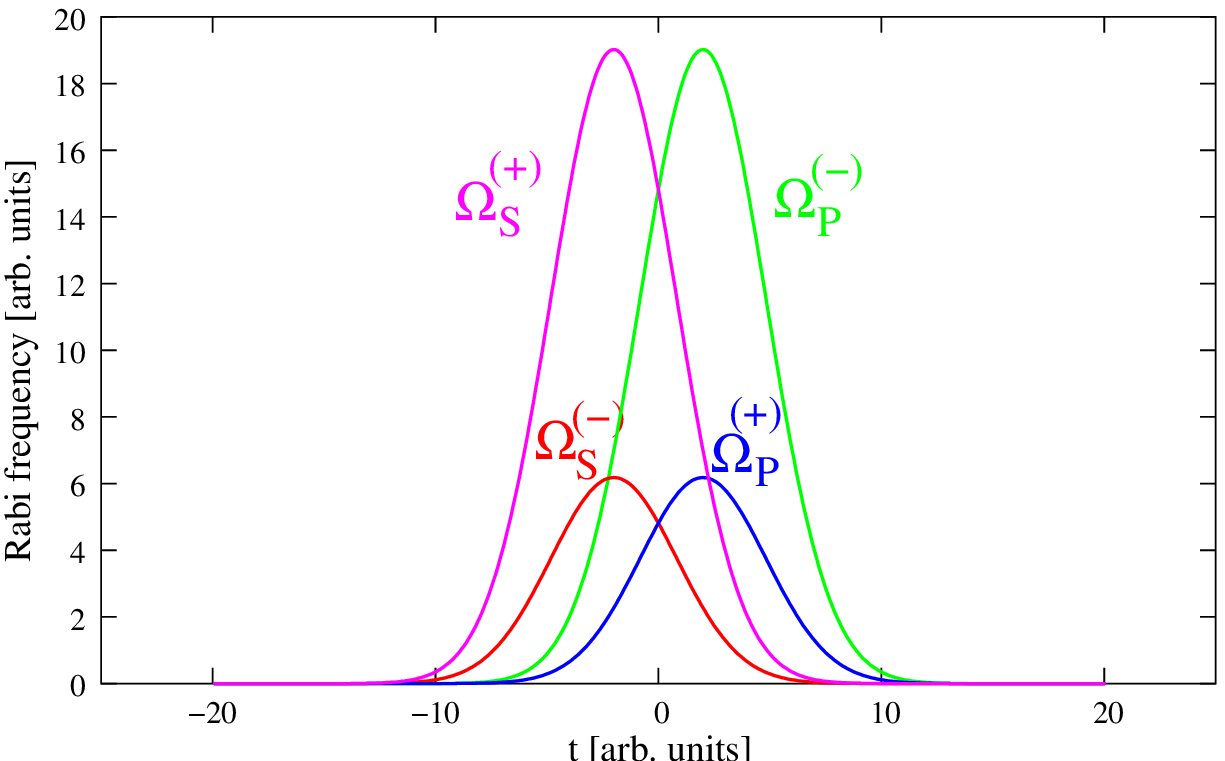}\\[10pt] 

\includegraphics[width=8cm]{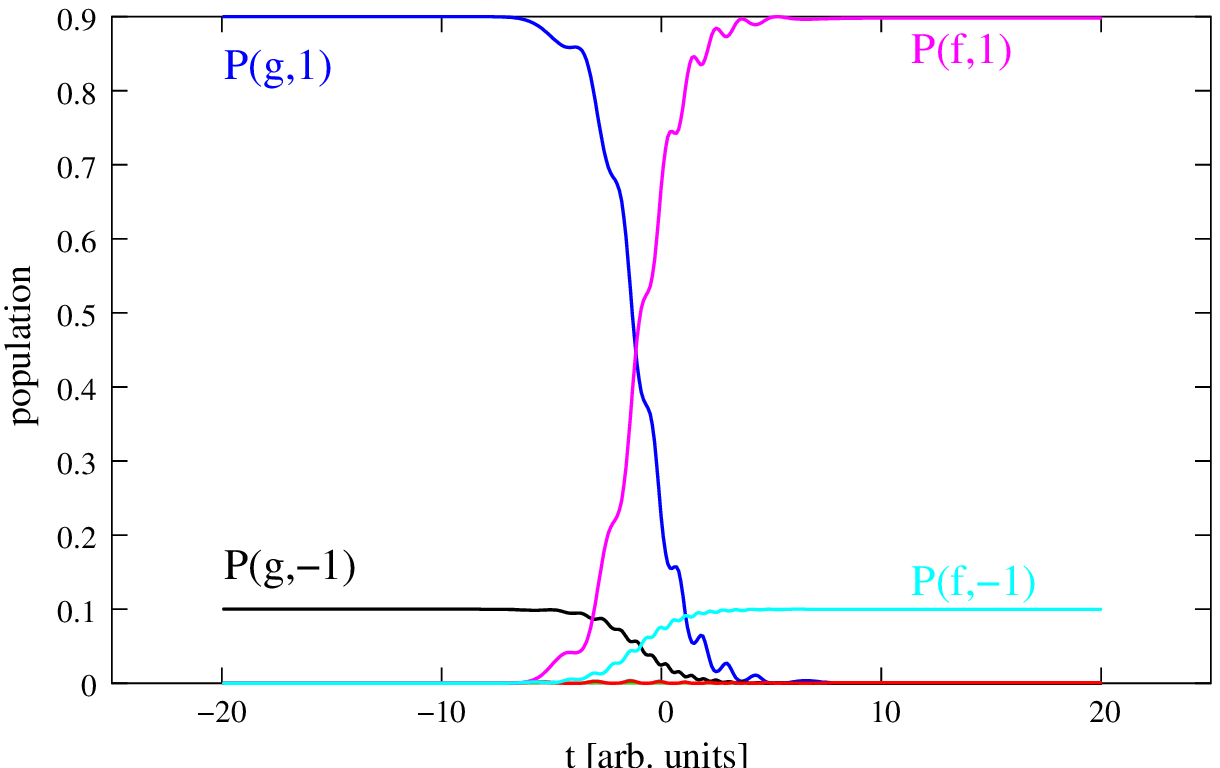} 
     \caption{ (Color Online) Same as Fig.~\ref{fig:121A-evol}, but now
       the    special   condition    for   the    pulse   polarizations
       Eq.~(\ref{cptdi2})  is   fulfilled.   Upper  frame:   the  pulse
       sequence  used   for  the  population   transfer  process.   The
       polarizations for  the pump and Stokes  pulses are $\eta=2\pi/5$
       and $\theta=\pi/10$, respectively.   Lower frame: The population
       evolution.  All population is  transferred from the $g$ set into
       the $f$ set. }
\label{fig:121B-evol}
\end{figure}

Fig.~\ref{fig:121A-evol} demonstrates  the population transfer  in the
twin  diamond configuration.  Initially  the system  was in  the state
$\cos(\alpha)           |g,J_g=1,M_g=-1\rangle+           \sin(\alpha)
|g,J_g=1,M_g=1\rangle$   with  $\alpha=\arctan(1/3)$.    The  envelope
functions  of   the  pump   and  Stokes  pulses   are  chosen   as  in
Sec.~\ref{subsec:111}.  The polarization of the pump and Stokes pulses
were chosen  as $\eta=2\pi/5$ and  $\theta=-\pi/7$, respectively.  All
phases of the  pulses are zero. The detuning $\Delta$ is  set to zero. 
After the  pulse sequence has passed,  some population is  left in the
$g$ and $e$  sets because the polarizations of  the pulses violate the
special condition for  complete transfer Eq.~(\ref{cptdi2}).  Finally,
in Fig.~\ref{fig:121B-evol}  the polarizations of the  pump and Stokes
pulses are chosen so  that the special condition Eq.~(\ref{cptdi2}) is
fulfilled.    Then,  a  complete   population  transfer   occurs,  all
population from the $g$ set is moved into the $f$ set.

\section{Summary}\label{sec:summary}

We have considered  the extension of the well-known  STIRAP process in
degenerate systems in which $N_g$ degenerate states of the $g$ set are
coupled by means of a pump pulse to $N_e$ degenerate states of the $e$
set, which in turn are linked  by the Stokes pulse to $N_f$ degenerate
states of the  $f$ set.  We have shown that  such a generalized STIRAP
process is always possible  if the succession of state-degeneracies is
nondecreasing,  i.e.  $  N_g \leq  N_e \leq  N_f$; and  the  number of
non-vanishing MS Rabi  frequencies is at least $N_g$  for both the pump
and Stokes couplings. When such conditions hold,  then for arbitrary
couplings  among states  (e.g.  arbitrary  elliptical  polarization of
electric dipole  radiation between magnetic sublevels)  it is possible
to obtain complete adiabatic passage of all population from the states
of the $g$ set into some combination of states of the $f$ set. In this
process the  initial state is arbitrary,  it can be any  pure or mixed
state that occupy the $g$ set.

An important exception  from the above rule occurs  in coupled angular
momentum systems,  when $J_g=J_e=J_f$.  Then,  due to the  symmetry of
the Clebsch-Gordan  coefficients some couplings  vanish, which results
in incomplete transfer.

We  have  examined the  possibility  of  adiabatic  passage when  this
restriction on degeneracies does not hold.  We have shown that part of
the  population can  be  transferred to  the  $f$ set.   We have  also
pointed  out that,  for certain  choices of  the polarizations  of the
coupling  fields,  complete   adiabatic  population  transfer  can  be
obtained.

We  have demonstrated  that  our scheme  can  be a  powerful tool  for
coherent control of  the quantum state in a  degenerate system: in our
proposal  the  selective  addressing   of  individual  states  in  the
degenerate sets is not required.  Nevertheless, the final state can be
tailored by varying  the polarizations and the relative  phases of the
coupling fields.   We have shown  through some specific  examples that
the control of the final superposition state is possible; the level of
control depends on the system under consideration.

\section*{Acknowledgments}

This work has  been supported by the European  Union Research Training
network  COCOMO,  contract   number  HPRN-CT-1999-00129.   ZK  and  AK
acknowledge  the  support from  the  Research  Fund  of the  Hungarian
Academy of Sciences (OTKA) under contract T43287.  ZK acknowledges the
support from  the J\'anos Bolyai  program of the Hungarian  Academy of
Sciences.   He is also  grateful to  Prof. K.   Bergmann for  his kind
hospitality in his group at  the University of Kaiserslautern. NVV and
BWS acknowledge  support from the  Alexander von Humboldt  Foundation.
BWS acknowledges support from the Graduierten Kolleg of the University
of Kaiserslautern.  The authors are  grateful to Prof. K. Bergmann for
useful discussion.

\appendix

\section{Dipole Transition Moments}

A common situation  where degeneracy occurs is when  the atomic states
are eigenstates of  angular momentum, bearing the labels  $J$ and $M$.
Then the  dipole moments can  be expressed in terms  of Clebsch-Gordan
coefficients  and reduced  matrix  elements. For  the pump  transition
($g-e$) the general pattern  of the dipole-transition matrix elements,
for arbitrary polarization, is
\begin{equation}
   \mu_{ij} = (g|\mu|e)   \sum_q \epsilon_q^{(p)}
   \frac{(J_g M_i, 1 q | J_e  M_j)}{  \sqrt{2 J_g + 1}}\,,
   \quad
   \left\{\begin{array}{l}
       i=1\hdots N_g \\
       j=1\hdots N_e
     \end{array}\right.\,,
\end{equation}
where $(g|\mu|e)$ is the reduced matrix element and $\epsilon_q^{(p)}$
parameterizes  the  contribution of  spherical  component  $q$ to  the
interaction.  The Stokes transition moments are similarly written as
\begin{equation}
  \mu_{ij} = (e|\mu|f)  \sum_q \epsilon_q^{(S)}\frac{(J_e M_i, 1 q | J_f M_j) }
  {\sqrt{2 J_e + 1}}\,,  \quad
  \left\{\begin{array}{l}
      i=1\hdots N_e \\
      j=1\hdots N_f
    \end{array}\right..
\end{equation}

For vibrational transitions in molecules the reduced matrix element
must include a Franck-Condon factor.

\section{Singular coupling matrix $\widetilde{\bm\Sigma}$}
\label{sec:sing-sigma}

Let  us consider  the  Hamiltonian Eq.~(\ref{ham})  in  the MS  basis,
Eq.~(\ref{trafo}). The MS transformation  of the coupling matrix ${\bm
  S}$ may result  in three different forms, shown  in Eq.~(\ref{Str}). 
We  obtain  a diagonal  matrix  $\widetilde{\bm\Sigma}$  to which  are
appended either rows  (if $N_f < N_e$) or columns (if  $N_f > N_e$) of
zero  values.  In  the discussions  of Sec.~\ref{sec:mstrafo}  we have
assumed that the  matrix $\widetilde{\bm\Sigma}$ is nonsingular.  Here
we   consider    the   case    when   some   diagonal    elements   of
$\widetilde{\bm\Sigma}$ are zero.  Let us choose the MS transformation
matrices ${\bm  A}$ and  ${\bm B}$ in  Eq.~(\ref{Udef}) in such  a way
that the zero  diagonal elements appear in the  bottom right corner of
$\widetilde{\bm\Sigma}$.    This   non-zero   part   is   denoted   by
$\widetilde{\bm\Sigma}_C$.   Let  the  dimension  of  this  matrix  be
$N_C\times N_C$.  In this notation, instead of Eq.~(\ref{Str}) we have
\begin{equation}\label{app:Sdef2}
   \widetilde{\bm S} = \left[\begin{array}{cc}
       \widetilde{\bm \Sigma}_C & {\bm 0} \\
       {\bm 0} & {\bm 0}
     \end{array}\right]\,,
\end{equation}
where the number  of all zero rows is $N_e-N_C$ and  the number of all
zero  columns is  $N_f-N_C$. From  this  form of  the Stokes  coupling
matrix it is clearly seen  that we have $N_e-N_C$ uncoupled MS states
in the $e$  set and $N_f-N_C$ uncoupled MS states in  the $f$ set.  By
inserting   the  coupling   matrix   Eq.~(\ref{app:Sdef2})  into   the
transformed Hamiltonian  of Eq.~(\ref{trafo}) and  performing a second
MS transformation  as in Sec.~\ref{sec:diamond} among the  $g$ set and
the uncoupled MS states of the $e$ set we get
\begin{equation}\label{app:Hamc}
   \widehat{\widetilde{{\bm H}}}(t)=\left[\begin{array}{cccccc}
       {\bm 0}&{\bm 0}&p(t)\widetilde{{\bm P}}&{\bm 0}&{\bm 0}&{\bm 0}\\
       {\bm 0}&{\bm 0}&p(t)\widetilde{{\bm
           P}}'&p(t)\widetilde{{\bm\Pi}}&{\bm 0}&{\bm 0}\\
       p(t)\widetilde{{\bm P}}^\dagger&p(t)\widetilde{{\bm P}}^{\prime\dagger}
       &\hbar{\bm \Delta}&{\bm 0}&s(t)\widetilde{\bm \Sigma}_C&{\bm 0}\\
       {\bm 0}&p(t)\widetilde{{\bm\Pi}}^{\dag}&{\bm 0}&\hbar{\bm \Delta}&{\bm 0}
       &{\bm 0}\\
       {\bm 0}&{\bm 0}&s(t)\widetilde{\bm \Sigma}_C^\dagger&{\bm
         0}&{\bm 0}&{\bm 0}\\
       {\bm 0}&{\bm 0}&{\bm 0}&{\bm 0}&{\bm 0}&{\bm 0}
     \end{array}\right]\,.
\end{equation}
This   Hamiltonian    is   almost   identical   with    the   one   in
Eq.~(\ref{Hambt}). The  difference is that  here on the bottom  of the
matrix we  have some rows  of zero values  as well as some  columns of
zero  values  to  the  far   right.   The  adiabatic  states  of  this
Hamiltonian   can  be   found  as   in   Sec.~\ref{sec:diamond}.   The
eigenvectors are parameterized as
\begin{equation}
\label{app:vkparam3}
\widetilde{\advecb}_k = \left[\begin{array}{c}
        \widetilde{\bm x}_k\\
        \widetilde{\bm x}'_k\\
        \widetilde{\bm y}_k\\
        \widetilde{\bm y}'_k\\
        \widetilde{\bm z}_k\\
        \widetilde{\bm z}'_k\\
        \end{array}\right].
\end{equation}
The   eigenvalue  equation  yields   the  set   of  equations   as  in
Sec.~\ref{sec:diamond} plus one more equation for ${\bm z}'_k$
\begin{equation}\label{app:extraeq}
      {\bm 0}=\eigenv_k{\bm z}'_k\,.
\end{equation}
When looking for  the eigenstates belonging to the  eigenvalue zero we
set $\eigenv_0=0$ in Eq.~(\ref{app:extraeq}).  Since ${\bm z}'_0$ does
not appear  in the other equations,  its value is  determined from the
initial  condition  of  the  system.   Our usual  assumption  is  that
initially  only the  states of  the $g$  set are  occupied, therefore,
${\bm z}'_0={\bm  0}$. For  the eigenstates with  non-zero eigenvalues
the only way to satisfy Eq.~(\ref{app:extraeq}) is to set ${\bm z}'_k$
to a null  vector , ${\bm z}'_k={\bm 0}$.   The eigenstates associated
with    non-zero     eigenvalues    $\eigenv_k$    are     given    in
Sec.~\ref{sec:diamond}.

\section{Linearization of the couplings $g\leftrightarrow e \leftrightarrow f$}
\label{sec:linearized-couplings}

The  construction of  the  dark state  Eq.~(\ref{darkstates1}) can  be
understood as follows.  We introduce  three sets of states, defined in
the $g$, $e$, and $f$ sets, respectively
\begin{subequations}\label{sets}
   \begin{eqnarray}
     \mbox{$g$ set: } \widetilde{\widetilde \psi}_g^{(l)}&=&
     {\bm x}_0^{(l)}\,,\quad l=1\hdots N_g\,,\label{gset}\\
     \mbox{$e$ set: } \widetilde{\widetilde \psi}_e^{(l)}&=&
     \frac{1}{{\cal N}^{(l)}_e}\widetilde{{\bm P}}^\dagger
       {\bm x}_0^{(l)}\,,\quad l=1\hdots N_g\,, \label{eset}\\
       &+& \mbox{$N_e-N_g$ other linearly independent states} \nonumber \\
     \mbox{$f$ set: } \widetilde{\widetilde \psi}_f^{(l)}&=&
     \frac{1}{{\cal N}^{(l)}_f}\widetilde{\bm \Sigma}^{-1}
     \widetilde{{\bm P}}^\dagger   {\bm x}_0^{(l)} \,,
     \quad l=1\hdots N_g\,,\label{fset}\\
     &+& \mbox{$N_f-N_g$ other linearly independent states}\,. \nonumber
   \end{eqnarray}
\end{subequations}
The vectors ${\bm x}_0^{(l)}$  are orthonormal by construction; ${\cal
   N}^{(l)}_e$  and ${\cal  N}^{(l)}_f$  are appropriate  normalization
factors for the other components.  The  states in the $g$ and $f$ sets
of Eq.~(\ref{sets}) are orthonormal, but  the states in the $e$ set of
Eq.~(\ref{eset}), though linearly independent and providing a complete
set  of excited  states,  are not  orthogonal.   The dual  counterpart
\cite{dual} of the $e$ set of Eq.~(\ref{eset}) reads
\begin{eqnarray}
   \mbox{dual $e$ set: } \widehat{\widehat \psi}_e^{(l)}&=&
   \frac{ {\cal N}^{(l)}_e}{{\cal N}^{(l)\,2}_f}
   {\bm x}_0^{(l)\,T}\widetilde{\bm P}  \widetilde{\bm \Sigma}^{-1 \dag}
   \widetilde{\bm \Sigma}^{-1}   \,,\\
   &&l=1\hdots N_g\,,\nonumber \\
   &+& \mbox{$N_e-N_g$ other linearly} \nonumber \\
   &&\mbox{independent states.} \nonumber
   \label{deset}
\end{eqnarray}
The vectors of these two sets are mutually orthogonal
\begin{equation}
   \langle \widehat{\widehat \psi}_e^{(l)}| \widetilde{\widetilde \psi}_e^{(k)}
   \rangle=\delta_{kl}\,.
\end{equation}
In  the  basis  defined  by  Eqs.   (\ref{sets})  the  Hamiltonian  of
Eq.~(\ref{ham-ms}) reads
\begin{equation}\label{ham-northo}
   \widetilde{\widetilde{{\bm H}}}(t)=\left[\begin{array}{cccc}
       {\bm 0}&p(t)\widetilde{{\bm Q}}_1&{\bm 0}&{\bm 0} \\
       p(t)\widetilde{{\bm Q}}_2&\hbar{\bm \Delta} &
       s(t) \widetilde{\bm \Sigma}_1 & {\bm 0} \\
       {\bm 0}&s(t)\widetilde{\bm \Sigma}_2 & {\bm 0}&{\bm 0}\\
       {\bm 0}&{\bm 0}&{\bm 0}&{\bm 0}
     \end{array}\right]\,,
\end{equation}
where  $\widetilde{{\bm Q}}_2$  and  $\widetilde{{\bm \Sigma}}_1$  are
diagonal matrices with elements
\begin{subequations}
   \begin{eqnarray}
     (\widetilde{{\bm  Q}}_{2})_{ll}&=&\frac{{\cal N}_f^{(l) 2}}
     {{\cal N}_e^{(l) }}\,,\quad l=1\hdots N_g\,,\\
     &\mbox{and}&\nonumber \\
     (\widetilde{{\bm  \Sigma}}_{1})_{ll}&=&\frac{1} {{\cal N}_f^{(l) }}
     (\widetilde{{\bm  Q}}_{2})_{ll}\,, \quad l=1\hdots N_g\,,
   \end{eqnarray}
\end{subequations}
respectively.   It  can  be  verified  that  the  matrix  elements  of
$\widetilde{\bm P}^{\dag}$  is zero between  the rest of the  dual $e$
states and the $g$ states
\begin{equation}
   \langle \widehat{\widehat \psi}_e^{(k)}|\widetilde{\bm P}^{\dag}|
   \widetilde{\widetilde \psi}_g^{(l)}  \rangle=0\,,\quad
   k=N_g+1\hdots N_e\,,\,\, l=1\hdots N_g\,.
\end{equation}
Similarly,  the matrix  elements of  $\widetilde{\bm \Sigma}$  is zero
between the rest of the dual $e$ states and the first $N_g$ $f$ states
\begin{equation}
   \langle \widehat{\widehat \psi}_e^{(k)}|\widetilde{\bm \Sigma}|
   \widetilde{\widetilde \psi}_f^{(l)}  \rangle=0\,,\quad
   k=N_g+1\hdots N_e\,,\,\, l=1\hdots N_g\,.
\end{equation}
The matrix elements of  the other two symmetric, non-diagonal matrices
$\widetilde{{\bm Q}}_1$ and $\widetilde{{\bm \Sigma}}_2$ are given by
\begin{subequations}
   \begin{eqnarray}
     (\widetilde{{\bm  Q}}_{1})_{lk}&=&\frac{1} {{\cal N}_e^{(k) }}
     \langle {\bm x}_0^{(l)\,T}|\widetilde{\bm P} \widetilde{\bm P}^{\dag}
     |{\bm x}_0^{(k)}\rangle\,,
     \\
     &\mbox{and}&\nonumber \\
     (\widetilde{{\bm  \Sigma}}_{2})_{lk}&=&\frac{1} {{\cal N}_f^{(l) }}
     (\widetilde{{\bm  Q}}_{1})_{lk}\,.
   \end{eqnarray}
\end{subequations}
The dark states of  the Hamiltonian (\ref{ham-northo}) can be obtained
in the  same manner as  in the above  derivation that led to  the dark
states Eq.~(\ref{darkstates1}).  The  population transfer is described
by the equation
\begin{equation}
   p(t){\cal N}_f^{(l) }\widetilde{x}_0^{(l)} +
   s(t)\widetilde{\widetilde{ z}}_0^{(l)}=0\,,
\end{equation}
where       the      components       $\widetilde{x}_0^{(l)}$      and
$\widetilde{\widetilde{z}}_0^{(l)}$  are  the  probability  amplitudes
associated with the basis  vectors Eqs.  (\ref{gset}) and (\ref{fset})
in  the $g$  and  $f$ sets,  respectively.   Hence in  this basis  the
couplings   $g\leftrightarrow  e   \leftrightarrow  f$   provide  {\em
   independent} pathways  of excitation.   Each $g$ state  is connected
through a single pathway to a single $f$ state.

\begin{widetext}

\section{Stokes field MS transformation matrices for the
$J=1\leftrightarrow2\leftrightarrow3$ linkage}\label{sec:123MS}

The   Stokes  field   MS  transformation   yields   three  eigenvalues
$\lambda_k$ of  the matrix ${\bm  S}{\bm S}^{\dag}$ composed  from the
Stokes field coupling matrix of Eq.~(\ref{S234})
\begin{equation}\label{eigvals123MS}
  \lambda_k = z+w\cot\left(\frac{1-k}{3}\pi+\frac{1}{3}\arctan v\right),
\end{equation}
for $k=1,2,3$, where
\begin{subequations}
  \begin{eqnarray}
    u &=& \frac{3}{4}\sqrt{146004\cos 12\theta + 857454 \cos 8\theta
      + 2234532\cos 4\theta + 1524810}\\
    v &=& \frac{2u}{(839+909\cos 4\theta)}\\
    w &=& \frac{73002\cos 12\theta+428727\cos 8\theta
      +1117266\cos 4\theta+762405}{22960u + 19320u\cos 4\theta}\\
    z &=& \frac{709 \cos 4\theta + 923}{14490\cos 4\theta+17220}.
  \end{eqnarray}
\end{subequations}
The Stokes field MS transformation matrix $\bm A$ is given by
\begin{equation}\label{A234}
  \bm A=e^{i\phi_S}\left[\begin{array}{cccc}
      
      p_1^{(A)}(\lambda_1)/n^{(A)}(\lambda_1)&p_2^{(A)}(\lambda_1)/n^{(A)}(\lambda_1)&p_3^{(A)}(\lambda_1)/n^{(A)}(\lambda_1)&p_4^{(A)}(\lambda_1)/n^{(A)}(\lambda_1)\\
      
      p_1^{(A)}(\lambda_2)/n^{(A)}(\lambda_2)&p_2^{(A)}(\lambda_2)/n^{(A)}(\lambda_2)&p_3^{(A)}(\lambda_2)/n^{(A)}(\lambda_2)&p_4^{(A)}(\lambda_2)/n^{(A)}(\lambda_2)\\
      
      p_1^{(A)}(\lambda_3)/n^{(A)}(\lambda_3)&p_2^{(A)}(\lambda_3)/n^{(A)}(\lambda_3)&p_3^{(A)}(\lambda_3)/n^{(A)}(\lambda_3)&p_4^{(A)}(\lambda_3)/n^{(A)}(\lambda_3)\\
      
      d_1^{(A)}/n_d^{(A)}&d_2^{(A)}/n_d^{(A)}&d_3^{(A)}/n_d^{(A)}&d_4^{(A)}/n_d^{(A)}
    \end{array}\right],
\end{equation}
where the polynomials $p_i^{(A)}(x)$ and the normalization $n_d^{(A)}(x)$ read
\begin{subequations}
  \begin{eqnarray}
    p_1^{(A)}(x)&=&\frac18e^{i(2\psi_S-2\phi_S)}\sin\theta(14700x^2-980(2+\cos
    2\theta)x+\cos4\theta+56\cos2\theta+63)\,,\\
    p_2^{(A)}(x)&=&\frac{\sqrt{15}}{24}e^{i(\psi_S-\phi_S)}
    \cos\theta(2940x^2-(308+280\cos  2\theta)x+3\cos4\theta+12\cos
    2\theta+9)\,,\\
    p_3^{(A)}(x)&=&\frac{\sqrt{15}}{8}\sin\theta(28(1+\cos2\theta)x-
    \cos4\theta-4\cos2\theta-3)\,,\\
    p_4^{(A)}(x)&=&\frac{1}{8}e^{i(\phi_S-\psi_S)}\cos\theta(1-\cos
    4\theta)\,,\\
    n^{(A)}(x)&=&\sqrt{ \left|p_1^{(A)}(x)\right|^2
      + \left|p_2^{(A)}(x)\right|^2 + \left|p_3^{(A)}(x)\right|^2 +
      \left|p_4^{(A)}(x)\right|^2}\,,
  \end{eqnarray}
  and the coefficients $d_i^{(A)}$, and the normalization $n_d^{(A)}$ are
  defined as
  \begin{eqnarray}
    d_1^{(A)}&=&-e^{i(2\psi_S-2\phi_S)}\cot^3\theta\,,\\
    d_2^{(A)}&=&\sqrt{15}e^{i(\psi_S-\phi_S)}\cot^2\theta\,,\\
    d_3^{(A)}&=&-\sqrt{15}\cot\theta\,,\\
    d_4^{(A)}&=&e^{i(\phi_S-\psi_S)}\,,\\
    n_d^{(A)}&=&\sqrt{1+15\cot^2\theta+15\cot^4\theta+\cot^6\theta}\,.
  \end{eqnarray}
\end{subequations}
Similarly, the other Stokes field MS transformation matrix is obtained
as
\begin{equation}\label{B234}
   \bm B=\left[\begin{array}{ccc}
       p_1^{(B)}(\lambda_1)/n^{(B)}(\lambda_1)&p_2^{(B)}(\lambda_1)/n^{(B)}(\lambda_1)&p_3^{(B)}(\lambda_1)/n^{(B)}(\lambda_1)\\
       p_1^{(B)}(\lambda_2)/n^{(B)}(\lambda_2)&p_2^{(B)}(\lambda_2)/n^{(B)}(\lambda_2)&p_3^{(B)}(\lambda_2)/n^{(B)}(\lambda_2)\\
       p_1^{(B)}(\lambda_3)/n^{(B)}(\lambda_3)&p_2^{(B)}(\lambda_3)/n^{(B)}(\lambda_3)&p_3^{(B)}(\lambda_3)/n^{(B)}(\lambda_3)
     \end{array}\right],
\end{equation}
where   the   polynomials   $p_i^{(B)}(x)$   and   the   normalization
$n_d^{(B)}(x)$ read
\begin{subequations}
  \begin{eqnarray}
    p_1^{(B)}(x)&=&e^{i(\phi_S-\psi_S)}p_1^{(A)}(x)/\sin\theta\,,\\
    p_2^{(B)}(x)&=&\frac{\sqrt{6}}{12}\sin 2\theta
    (105 x -7 \cos 2\theta -8)\,,\\
    p_3^{(B)}(x)&=&\frac{1}{4}e^{i(\phi_S-\psi_S)}\sin^2 2\theta\,,\\
    n^{(B)}(x)&=&\sqrt{ \left|p_1^{(B)}(x)\right|^2
      +\left|p_2^{(B)}(x)\right|^2 +\left|p_3^{(B)}(x)\right|^2}\,.
  \end{eqnarray}
\end{subequations}
The  vectors ${\bm  x}_0^{(1,2)}$  characterizing the  dark states  of
Eq.~(\ref{darkstates1})  are obtained by  finding the  eigenvectors of
the  Hermitian   matrix  Eq.~(\ref{metric}),  which   is  obtained  by
inserting  Eqs.~(\ref{P234}), (\ref{sigma234})  and  (\ref{B234}) into
Eq.~(\ref{metric}). The two eigenvectors are given by
\begin{subequations}\label{x0:123}
  \begin{eqnarray}
    {\bm x}_0^{(1)}
    &=&\left[\begin{array}{c}
        \sin\chi e^{i\xi}\\
        \cos\chi
      \end{array}\right],\\
    {\bm x}_0^{(2)} &=&\left[\begin{array}{c}
        \cos\chi e^{i\xi}\\
        -\sin\chi
      \end{array}\right],
  \end{eqnarray}
\end{subequations}
where
\begin{subequations}
  \begin{eqnarray}
    \chi&=&\frac{1}{2}\arctan\frac{2|u'|}{v'},
    \\
    \xi&=&\arg u',
    \\
    u'&=&\frac{7}{60}e^{i(\phi_S-\psi_S)}\sin2
    \theta(-8+7\cos2\theta\cos2\eta) \nonumber\\
    &&\quad +e^{i(\phi_P-\psi_P)}\sin2\eta\left[\frac{7}{24}
      +\left(\frac{343}{360}+\frac{7}{40}e^{2i(\phi_S-\psi_S+
          \psi_P-\phi_P)}\right) \sin^22\theta\right],
    \\
    v'&=&\frac{49}{60}\cos(\phi_S-\psi_S+\psi_P-\phi_P)
    \sin2\eta\sin4\theta  +\left(\frac{301}{36}+
      \frac{203}{90}\cos^22\theta\right)\cos2\eta-\frac{49}{5}\cos2\theta.
  \end{eqnarray}
\end{subequations}

\section{Stokes field MS transformation matrices for the
$J=1\leftrightarrow2\leftrightarrow1$ linkage}\label{sec:121MS}

The Stokes field MS transformation matrix $\bm A$ is given by
\begin{equation}
   \bm A = e^{i\psi_S}\left[\begin{array}{cc}
       p_1^{(A)}(\lambda_1)/n^{(A)}(\lambda_1)&p_2^{(A)}(\lambda_1)/
       n^{(A)}(\lambda_1)\\
       p_1^{(A)}(\lambda_2)/n^{(A)}(\lambda_2)&p_2^{(A)}(\lambda_2)/
       n^{(A)}(\lambda_2)
     \end{array}\right]\,,
\end{equation}
where the polynomials $p_i^{(A)}(x)$ and the normalization $n^{(A)}(x)$ read
\begin{eqnarray}
   p_1^{(A)}(x)&=&-1-5\sin^2\theta+50 x,\\
   p_2^{(A)}(x)&=&\sin\theta\cos\theta e^{-i(\psi_S-\phi_S)},\\
   n^{(A)}(x)&=&\sqrt{ \left|p_1^{(A)}(x)\right|^2 +
     \left|p_2^{(A)}(x)\right|^2}.
\end{eqnarray}
Similarly, the other Stokes field MS transformation matrix is obtained
as
\begin{equation}
   \bm B = \left[\begin{array}{c}
       \bm B_a\\
       \bm B_b
     \end{array}\right]\,,
\end{equation}
where
\begin{equation}
   \bm B_a = \left[\begin{array}{cc}
       \mathrm{sgn}(\sin4\theta\sin\theta)&0\\
       0&\mathrm{sgn}(\cos \theta)
     \end{array}\right]
   \left[\begin{array}{ccc}
       p_1^{(B)}(\lambda_1)/n^{(B)}(\lambda_1)&p_2^{(B)}
       (\lambda_1)/n^{(B)}(\lambda_1)&p_3^{(B)}(\lambda_1)/n^{(B)}(\lambda_1)\\
       p_1^{(B)}(\lambda_2)/n^{(B)}(\lambda_2)&p_2^{(B)}(\lambda_2)/n^{(B)}
       (\lambda_2)&p_3^{(B)}(\lambda_2)/n^{(B)}(\lambda_2)
     \end{array}\right]\,,
\end{equation}
and
\begin{equation}
   \bm B_b = \left[\begin{array}{ccc}
       d_1^{(B)}/n_d^{(B)}&d_2^{(B)}/n_d^{(B)}&d_3^{(B)}/n_d^{(B)}
     \end{array}\right]\,.
\end{equation}
The polynomials $p_i^{(B)}(x)$ with  the normalization $n_d^{(B)}(x)$;
the coefficients $d_i^{(B)}$ with the normalization $n_d^{(B)}$ read
\begin{eqnarray}
   p_1^{(B)}(x)&=&-e^{-i(\phi_S-\psi_S)} \cos^2\theta
   (7\sin^2\theta+\cos^2\theta-50x)\,,\\
   p_2^{(B)}(x)&=&\frac{\sqrt{6}}{4}\sin4\theta\,,\\
   p_3^{(B)}(x)&=&e^{-i(\psi_S-\phi_S)} \sin^2\theta
   (7\cos^2\theta+sin^2\theta-50x)\,,\\
   n^{(B)}(x)&=&\sqrt{ \left|p_1^{(A)}(x)\right|^2 +
     \left|p_2^{(A)}(x)\right|^2 + \left|p_3^{(A)}(x)\right|^2}\,,\\
   d_1^{(B)}&=&e^{i(\phi_S-\psi_S)}\sin^2\theta\,,\\
   d_2^{(B)}&=&-\sqrt{6}\sin\theta\cos\theta\,,\\
   d_3^{(B)}&=&e^{i(\psi_S-\phi_S)}\cos^2\theta\,,\\
   n_d^{(B)}&=&\sqrt{1+\sin^2 2\theta}\,.
\end{eqnarray}

\end{widetext}


\begin{thebibliography}{99}

\bibitem{Vitanov01} N.V. Vitanov, M.  Fleischhauer, B.W. Shore, and K.
   Bergmann, Adv.  Atomic Mol. Opt. Phys. {\bf 46}, 55 (2001).

\bibitem{STIRAP} J.  Oreg, F. T.  Hioe, J.H. Eberly, Phys. Rev. A {\bf
     29}, 690 (1984); J.  R.   Kuklinski, U.  Gaubatz, F.  T. Hioe, and
   K. Bergmann, Phys.   Rev.  A {\bf 40}, 6741 (1989);  U.  Gaubatz, P.
   Rudecki, S.  Schiemann, and K.   Bergmann, J. Chem.  Phys. {\bf 92},
   5363 (1990).

\bibitem{AAMOP}  N.V.  Vitanov,  T.  Halfmann,  B.W.  Shore,  and  K.
   Bergmann, Ann. Rev. Phys. Chem. {\bf 52}, 763 (2001)

\bibitem{Shore91} B.W. Shore, K.  Bergmann, J. Oreg, and S. Resenwaks,
   Phys. Rev. A {\bf 44}, 7442 (1991).

\bibitem{Smith92} A.V. Smith, J. Opt. Soc. Am. B {\bf 9}, 1543 (1992).

\bibitem{Pillet93}  P. Pillet, C.  Valentin, R.-L.   Yuan, and  J. Yu,
   Phys. Rev. A {\bf 48}, 845 (1993).

\bibitem{Weiss94} D.S. Weiss, B.C. Young,  S. Chu, Appl.  Phys. B {\bf
     59}, 217 (1994).

\bibitem{Shore95}  B.W. Shore,  J. Martin,  M.P. Fewell,  K. Bergmann,
   Phys. Rev. A {\bf 52}, 566 (1995).

\bibitem{Martin95} J. Martin, B. W. Shore, and K. Bergmann, Phys. Rev.
   A {\bf 52}, 583 (1995).

\bibitem{Malinovsky97}  V.S.  Malinovsky,  D.J. Tannor,  Phys.  Rev. A
   {\bf 56}, 4929 (1997).

\bibitem{Vitanov98} N.V. Vitanov, Phys. Rev. A {\bf 58}, 2295 (1998).

\bibitem{Theuer98} H.  Theuer and K.  Bergmann, Eur.  Phys. J.  D {\bf
     2}, 279 (1998).

\bibitem{Marte91} P.~Marte, P.~Zoller and J.L.~Hall, Phys. Rev. A {\bf
     44}, R4118 (1991).

\bibitem{Lawall94} J.  Lawall and M.  Prentiss, Phys. Rev.  Lett. {\bf
     72}, 993 (1994).

\bibitem{Goldner94}  L.S.  Goldner,  C.  Gerz, R.J.C.  Spreeuw,  S.L.
   Rolston, C.I.  Westbrook, W.D. Phillips,  P. Marte, and  P.  Zoller,
   Phys. Rev. Lett. {\bf 72}, 997 (1994).

\bibitem{Weitz94} M. Weitz,  B.C. Young, and S. Chu,  Phys. Rev. Lett.
   {\bf 73} 2563 (1994).

\bibitem{Unanyan98} R.G. Unanyan, M.  Fleischhauer, B.W. Shore, and K.
   Bergmann, Opt. Commun. {\bf 155}, 144 (1998).

\bibitem{Theuer99} H.  Theuer, R.G.  Unanyan, C.  Habscheid, K. Klein,
   and K. Bergmann, Opt. Express {\bf 4}, 77 (1999).

\bibitem{Unanyan99} R.G.  Unanyan, B.W. Shore, and K.  Bergmann, Phys.
   Rev. A {\bf 59}, 2910 (1999).

\bibitem{Unanyan01} R.G.  Unanyan, B.W. Shore, and K.  Bergmann, Phys.
   Rev. A {\bf 63}, 043401 (2001).

\bibitem{Kis01} Z. Kis and S.  Stenholm, Phys. Rev. A {\bf 64}, 063406
   (2001).

\bibitem{Kis02} Z. Kis  and S. Stenholm, J. Mod.  Optics {\bf 49}, 111
   (2002).

\bibitem{Kraal02a} P. Kr\'al, Z. Amitay, M. Shapiro, Phys.  Rev. Lett.
   {\bf 89}, 63002 (2002).

\bibitem{Kraal02b} P. Kr\'al, M. Shapiro, Phys. Rev. A {\bf 65}, 43413
   (2002).

\bibitem{Kis03}  A.  Karpati  and Z.  Kis, J.  Phys. B  {\bf  36}, 905
   (2003).

\bibitem{Morris83} J.  R. Morris and B. W.  Shore, Phys.  Rev.  A {\bf
     27}, 906 (1983).

\bibitem{5ss} N.V. Vitanov, Z. Kis, and B. W. Shore, Phys. Rev. A {\bf
     68}, 063414 (2003).


\bibitem{dual}  S. Lipschutz,  M.L. Lipson,  {\em Schaum's  Outline of
     Linear Algebra} (McGraw-Hill).

\bibitem{Messiah}  A. Messiah,  1959,  {\em  M\'ecanique  Quantique}
   (Paris: Dunod) pp.637-650.

\bibitem{Shah02} S.P. Shah, D.J.  Tannor,  and S.A. Rice, Phys. Rev. A
   {\bf 66}, 033405 (2002).





\end{thebibliography}
\end{document}